\documentclass[prd,superscriptaddress,preprintnumbers,twocolumn,showpacs,amsmath,amssymb]{revtex4-1}
\usepackage{graphicx}
\usepackage{color}
\usepackage{booktabs}
\usepackage{hyperref}
\usepackage{physics,slashed,braket}

\begin{document}
  \title{Berezinskii-Kosterlitz-Thouless transition in lattice Schwinger model with one flavor of 
  Wilson fermion}

\preprint{UTHEP-711, UTCCS-P-109}

  \author{Yuya Shimizu}
  \affiliation{RIKEN Advanced Institute for Computational Science, Kobe, Hyogo 650-0047, Japan}

  \author{Yoshinobu Kuramashi}
  \affiliation{Center for Computational Sciences, University of Tsukuba, Tsukuba, Ibaraki
    305-8577, Japan}
  \affiliation{RIKEN Advanced Institute for Computational Science, Kobe, Hyogo 650-0047, Japan}

  \begin{abstract}
    We have made a detailed study of the phase structure for lattice Schwinger model with one flavor of Wilson fermion on the $(m,g)$ plane. For numerical investigation, we develop a decorated tensor
    renormalization method for lattice gauge theories with fermions incorporating the Grassmann
    tensor renormalization. Our algorithm manifestly preserves rotation and reflection symmetries.
    We find not only a parity-broken phase but also a Berezinskii-Kosterlitz-Thouless (BKT) transition
    by evaluating the central charge and an expectation value of a projection operator into the parity-odd subspace. The BKT phase boundaries converge into the degenerated doubler pole $(m,g)=(-2,0)$, while the
    parity-breaking transition line ends at the physical pole $(m,g)=(0,0)$. In addition, our analysis of
    scaling dimensions indicates that a conformal field theory with $\mathrm{SU}(2)$ symmetry arises on the
    line of $m=-2$.
  \end{abstract}
  \date{\today}
  \maketitle

  \section{Introduction}
  \label{sec:int}
  The Wilson fermion is one of the standard lattice regularizations for the continuum Dirac fermion,
  avoiding the species-doubling problem. Although the chiral symmetry is explicitly broken in this
  formulation, the existence of massless pseudoscalar mesons is expected by Aoki's
  scenario for even numbers of flavors as gapless modes accompanied by a parity-flavor-breaking transition
  (see a summary report~\cite{Aoki:1987us}). On the other hand, the phase structure of the Wilson fermion with odd numbers of flavors is not fully understood yet. The condition of the Vafa-Witten
  theorem~\cite{Vafa:1984xg} is not fulfilled in this case and the parity symmetry itself can be spontaneously broken,
  that is, the flavor singlet pseudoscalar meson can be also massless. In fact, analyses based on the strong-coupling
  expansion show that the parity symmetry is really broken. A possible answer was indicated by numerical simulations for three flavor lattice QCD: The redundant phase is restricted within the strong-coupling region and
  the correct continuum physics can be realized in a weak-coupling region~\cite{Aoki:2004iq}.
  However, it is still quite hard to investigate the parity- or parity-flavor-breaking transition,
  because Monte Carlo approaches suffer from the numerical sign problem in the broken phase for odd numbers of flavors.

  As a first step toward a long-term goal to understand the phase structure of odd numbers of flavors of the Wilson
  fermion completely, we study the lattice Schwinger model with one flavor of the Wilson fermion.
  The Schwinger model, two dimensional QED, is simple but shares several properties with QCD. It was
  numerically verified by several methods at high level of precision that an Ising transition occurs in the
  strong-coupling limit~\cite{Gausterer:1995np,Wenger:2008tq,Shimizu:2014uva}. Moreover, we showed that
  such transition is retained even at finite couplings and the transition line approaches the physical pole
  $(m,g)=(0,0)$ where $m$ is the fermion mass and $g$ is the gauge coupling~\cite{Shimizu:2014uva}.  On the other hand, the situation of the lattice Schwinger model around $m=-2$ is not clear yet.
In the strong coupling limit of $g=\infty$, as discussed by Salmhofer, it is mapped to a
  critical six-vertex model whose criticality is described by a conformal field theory (CFT) with the central
  charge $c=1$~\cite{Salmhofer:1991cc}. In the free limit, the point of $(m,g)=(-2,0)$ corresponds to the degenerated doubler pole, which is governed by the CFT with $c=\frac{1}{2}+\frac{1}{2}=1$. A natural expectation may be an existence of a $c=1$ CFT line
  or region between $(m,g)=(-2,0)$ and $(m,g)=(-2,\infty)$, which should be confirmed by a detailed investigation of the phase structure around $m=-2$ at finite couplings by using a reliable numerical method.

  In order to investigate the phase structure of the lattice Schwinger model with one flavor of the Wilson
  fermion throughout the $(m,g)$ plane, we have improved our method proposed in Ref.~\cite{Shimizu:2014uva} which
  was based on the tensor renormalization group (TRG)~\cite{Levin:2007aa}. The TRG is one of the so-called
  tensor network algorithms which are free from the sign problem and is suitable for computation of classical
  partition functions and quantum path-integrals. It was originally developed mainly in the fields of condensed
  matter physics and quantum information science. In recent years, however, it also attracts a lot of attention in
  high energy physics and there have been already many applications to lattice field
  theories~\cite{Shimizu:2012zza,Liu:2013nsa,Denbleyker:2013bea,Zou:2014rha,Shimizu:2014uva,Shimizu:2014fsa,
  Dittrich:2014mxa,Takeda:2014vwa,Bazavov:2015kka,Kawauchi:2016xng,Bazavov:2017hzi,Sakai:2017jwp,
  Yoshimura:2017jpk,Delcamp:2016dqo}.
  Although our previous method succeeded in clear demonstration of the Ising transition at finite
  couplings by adopting a Grassmann version of the TRG proposed by Gu and others~\cite{Gu:2010aa},
  the gauge redundancy demanded increasing bond dimensions of tensor to keep numerical precisions at finite couplings. After our work of Refs.~\cite{Shimizu:2014uva,Shimizu:2014fsa}, Dittrich and others proposed a
  variation of the TRG method which gives a better control of the gauge redundancy by introducing an additional structure decorating
  the tensor network~\cite{Dittrich:2014mxa}. In this paper we develop a more efficient method to study the
  lattice Schwinger model with one flavor of the Wilson fermion by combining the ideas of the decorated TRG and Grassmann TRG, and investigate the
  phase structure in the region around $m=-2$ at finite couplings, which was out of reach in previous studies. 

It is worth noting
  that other tensor network methods are also gathering much attention from the field of high energy physics. They also choose the analysis of the lattice Schwinger model as a pilot study, though the Kogut-Susskind fermion formulation has been employed so far in all such works~\cite{Byrnes:2002nv,Banuls:2013jaa,Rico:2013qya,Buyens:2013yza,Kuhn:2014rha,Banuls:2015sta,
  Pichler:2015yqa,Buyens:2015tea,Banuls:2016lkq,Buyens:2016ecr,Banuls:2016gid,Buyens:2017crb,Zapp:2017fcr,
  Ercolessi:2017jbi}.

  The layout of this papers is as follows. In Sec.~\ref{sec:met}, we explain the application of our decorated
  tensor renormalization to the lattice Schwinger model with one flavor of the Wilson fermion.
  Section~\ref{sec:num} presents the results for numerical investigation of the phase structure.
  Finally we summarize our paper in Sec.~\ref{sec:sum}.

  \section{Decorated tensor renormalization for the lattice Schwinger model}
  \label{sec:met}
  \subsection{Lattice Schwinger model with one flavor of the Wilson fermion}
  We follow the lattice formulation of the Schwinger model given in Ref.~\cite{Shimizu:2014uva}, but use
  $m$ and $g$ as parameters in the model throughout this paper instead of the hopping parameter $\kappa$
  and the inverse coupling squared $\beta$. We give the formulation again in the following to make the paper self-contained.

  The Wilson-Dirac matrix is given by
  \begin{equation}
    \begin{split}
      \Bar{\psi}D[U]\psi=&(m+2)\sum_{n,\alpha} \Bar{\psi}_{n,\alpha}\psi_{n,\alpha}\\
      &-\frac{1}{2}\sum_{n,\mu,\alpha,\beta}\Bar{\psi}_{n,\alpha}\bigl[
      (1-\gamma_\mu)_{\alpha,\beta}\,U_{n,\mu}\psi_{n+\Hat{\mu},\beta}\\
      &+(1+\gamma_\mu)_{\alpha,\beta}\,U^\dagger_{n-\Hat{\mu},\mu}\psi_{n-\Hat{\mu},\beta}\bigr],
    \end{split}
  \end{equation}
  where $\psi_n$ and $\Bar{\psi}_n$ are two-component Grassmann variables at site $n$ and $U_{n,\mu}$ is an
  $U(1)$ link variable at site $n$ along $\mu$ direction. $\alpha$ and $\beta$ denote the Dirac indices and
  $\Hat{\mu}$ represents an unit vector along $\mu$ direction. The gamma matrices are
  \begin{equation}
    \gamma_1=
    \begin{pmatrix}
      1&0\\
      0&-1
    \end{pmatrix},\quad
    \gamma_2=
    \begin{pmatrix}
      0&1\\
      1&0
    \end{pmatrix}.
  \end{equation}
  Following Ref.~\cite{Salmhofer:1991cc}, we employ another basis,
  \begin{gather}
    \chi_{n,1}=\frac{1}{\sqrt{2}}(\psi_{n,1}+\psi_{n,2}),\ 
    \chi_{n,2}=\frac{1}{\sqrt{2}}(\psi_{n,1}-\psi_{n,2}),\\
    \Bar{\chi}_{n,1}=\frac{1}{\sqrt{2}}(\Bar{\psi}_{n,1}+\Bar{\psi}_{n,2}),\ 
    \Bar{\chi}_{n,2}=\frac{1}{\sqrt{2}}(\Bar{\psi}_{n,1}-\Bar{\psi}_{n,2}),
  \end{gather}
  only in the second (space) direction, while keeping $\{\psi_{n,\alpha}\}$ and $\{\Bar{\psi}_{n,\alpha}\}$
  in the first (time) direction. Using this trick, the hopping terms are simply rewritten as
  \begin{gather}
    \sum_{\alpha,\beta}\Bar{\psi}_{n,\alpha}(1+\gamma_1)_{\alpha,\beta} \psi_{n-\Hat{1},\beta}
    =2\Bar{\psi}_{n,1}\psi_{n-\Hat{1},1},\\
    \sum_{\alpha,\beta}\Bar{\psi}_{n,\alpha}(1-\gamma_1)_{\alpha,\beta} \psi_{n+\Hat{1},\beta}
    =2\Bar{\psi}_{n,2}\psi_{n+\Hat{1},2},\\
    \sum_{\alpha,\beta}\Bar{\psi}_{n,\alpha}(1+\gamma_2)_{\alpha,\beta} \psi_{n-\Hat{2},\beta}
    =2\Bar{\chi}_{n,1}\chi_{n-\Hat{2},1},\\
    \sum_{\alpha,\beta}\Bar{\psi}_{n,\alpha}(1-\gamma_2)_{\alpha,\beta} \psi_{n+\Hat{2},\beta}
    =2\Bar{\chi}_{n,2}\chi_{n+\Hat{2},2}.
  \end{gather}
  Each Grassmann variable satisfies the anticommutation relations,
  \begin{gather}
    [\psi_{n,\alpha},\Bar{\psi}_{m,\beta}]_+\equiv \psi_{n,\alpha}\Bar{\psi}_{m,\beta}
    +\Bar{\psi}_{m,\beta}\psi_{n,\alpha}=0,\label{eq:ac1}\\
    [\psi_{n,\alpha},\psi_{m,\beta}]_+=[\Bar{\psi}_{n,\alpha},\Bar{\psi}_{m,\beta}]_+=0,\label{eq:ac2}\\
    [\chi_{n,\alpha},\Bar{\chi}_{m,\beta}]_+=[\chi_{n,\alpha},\chi_{m,\beta}]_+
    =[\Bar{\chi}_{n,\alpha},\Bar{\chi}_{m,\beta}]_+=0,\label{eq:ac3}
  \end{gather}
  and the integration rules,
  \begin{gather}
    \int\dd{\psi_{n,\alpha}}1=\int\dd{\Bar{\psi}_{n,\alpha}}1=0,\\
    \int\dd{\psi_{n,\alpha}}\psi_{m,\beta}=\int\dd{\Bar{\psi}_{n,\alpha}}\Bar{\psi}_{m,\beta}=\delta_{n,m}\,
    \delta_{\alpha,\beta},\\
    \int\dd{\chi_{n,\alpha}}1=\int\dd{\Bar{\chi}_{n,\alpha}}1=0,\\
    \int\dd{\chi_{n,\alpha}}\chi_{m,\beta}=\int\dd{\Bar{\chi}_{n,\alpha}}\Bar{\chi}_{m,\beta}=\delta_{n,m}\,
    \delta_{\alpha,\beta}.
  \end{gather}
  Turning to the gauge part, the $\mathrm{U}(1)$ plaquette action is given by
  \begin{gather}
    S_\mathrm{g}[U]=-\frac{1}{g^2}\sum_{\text{all plaquette}}\cos\varphi_\mathrm{p},\label{eq:ga}\\
    \varphi_\mathrm{p}\equiv\varphi_{n,1}+\varphi_{n+\Hat{1},2}-\varphi_{n+\Hat{2},1}-\varphi_{n,2},\\
    \varphi_{n,1},\varphi_{n+\Hat{1},2},\varphi_{n+\Hat{2},1},\varphi_{n,2}\in [-\pi,\pi],
  \end{gather}
  where $\varphi_{n,1},\varphi_{n+\Hat{1},2},\varphi_{n+\Hat{2},1}$ and $\varphi_{n,2}$ are phases of
  link variables which compose a plaquette variable $\varphi_\mathrm{p}$.

  \subsection{Decorated tensor network representation}
\begin{figure}
    \centering
    \includegraphics[width=60mm]{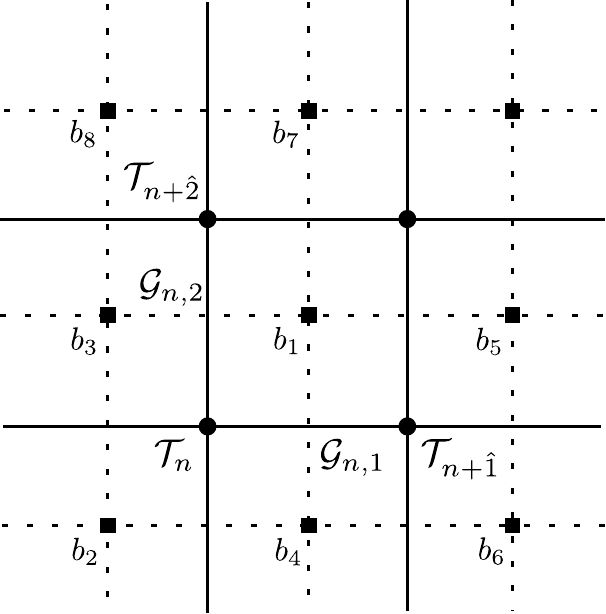}
    \caption{Schematic representation of the decorated tensor network of Eq.~\eqref{eq:tn}. Each site indicated
    by a filled circle represents each decorated tensor $\mathcal{T}_n$. $\{\mathcal{G}_{n,\mu}\}$ are
    assigned to each link. The decoration labels $\{b_i\}$ sit on each plaquette.}
    \label{fig:dtn}
  \end{figure}

  We transform the partition function,
  \begin{equation}
    Z=\int\mathcal{D}U\det D[U]\,e^{-S_\mathrm{g}[U]},
  \end{equation}
  into a decorated tensor network form. First, by using Eqs.~\eqref{eq:ac1}-\eqref{eq:ac3}, the fermionic
  part is expanded as follows:
  \begin{gather}
    \begin{align}
      \det D[U]&=\prod_{n,\alpha}\left(\int\dd{\psi_{n,\alpha}}\dd{\Bar{\psi}_{n,\alpha}}\right)
      e^{\Bar{\psi}D[U]\psi}\\
      &=P_0\int\prod_n W_n\prod_\mu H_{n,\mu},
    \end{align}\\
    \begin{split}
      W_n\equiv &(m+2)^2+(m+2)\dd{\psi_{n,1}}\dd{\Bar{\psi}_{n,1}}\\
      &+(m+2)\dd{\psi_{n,2}}\dd{\Bar{\psi}_{n,2}}
      +\dd{\psi_{n,1}}\dd{\Bar{\psi}_{n,1}}\dd{\psi_{n,2}}\dd{\Bar{\psi}_{n,2}},
    \end{split}\\
    \begin{split}
      H_{n,1}\equiv&1+U^\dagger_{n,1}\psi_{n,1}\Bar{\psi}_{n+\Hat{1},1}
      +U_{n,1}\psi_{n+\Hat{1},2}\Bar{\psi}_{n,2}\\&+\psi_{n+\Hat{1},2}\Bar{\psi}_{n+\Hat{1},1}
      \Bar{\psi}_{n,2}\psi_{n,1},
    \end{split}\\
    \begin{split}
      H_{n,2}\equiv&1+U^\dagger_{n,2}\chi_{n,1}\Bar{\chi}_{n+\Hat{2},1}
      +U_{n,2}\psi_{n+\Hat{2},2}\Bar{\chi}_{n,2}\\&+\Bar{\chi}_{n+\Hat{2},1}\chi_{n+\Hat{2},2}
      \chi_{n,1}\Bar{\chi}_{n,2},
    \end{split}
  \end{gather}
  where $P_0$ represents a projection to terms without any Grassmann variable.
  Next, we perform the character expansion for the gauge part:
  \begin{align}
    e^{\frac{1}{g^2}\cos\varphi_\mathrm{p}}&=\sum_{b_\mathrm{p}=-\infty}^\infty
    e^{ib_\mathrm{p}\varphi_\mathrm{p}}\,I_{b_\mathrm{p}}\pqty{\frac{1}{g^2}},\\
    &\simeq\sum_{b_\mathrm{p}=-N_\mathrm{ce}}^{N_\mathrm{ce}} e^{ib_\mathrm{p}\varphi_\mathrm{p}}\,
    I_{b_\mathrm{p}}\pqty{\frac{1}{g^2}},
  \end{align}
  with $I_{b_\mathrm{p}}$ the modified Bessel function. The truncation number $N_\mathrm{ce}$ is introduced
  in order to obtain a finite-dimensional tensor network~\cite{Liu:2013nsa}. Integrating out all link
  variables, we obtain
  \begin{multline}
    \int_{-\pi}^{\pi}\!\frac{d\varphi_{n,1}}{2\pi}H_{n,1}\,e^{i(b_i-b_j)\varphi_{n,1}}\\
    =
    \begin{cases}
      1+\psi_{n+\Hat{1},2}\Bar{\psi}_{n+\Hat{1},1}\Bar{\psi}_{n,2}\psi_{n,1},&b_i=b_j\\
      \psi_{n,1}\Bar{\psi}_{n+\Hat{1},1},&b_i=b_j+1\\
      \psi_{n+\Hat{1},2}\Bar{\psi}_{n,2},&b_i=b_j-1\\
      0,&\text{others}
    \end{cases}.
  \end{multline}
  Integration for $H_{n,2}$ is performed likewise. Multiplying $W_n$ by these integrated results yields a
  decorated tensor network form of the partition function:
  \begin{gather}
    \begin{split}
      Z=&\int\sum_{b_1,b_2,b_3,\cdots}\sum_{i,j,k,\cdots}\mathcal{T}^{(b_1,b_2,b_3,b_4)}_{n;i,j,k,l}\,
      \mathcal{T}^{(b_5,b_4,b_1,b_6)}_{n+\Hat{1};m,o,i^\prime,p}\\
      &\times \mathcal{T}^{(b_7,b_3,b_8,b_1)}_{n+\Hat{2};q,r,s,j^\prime}\,\cdots\,
      \mathcal{G}^{(b_1,b_4)}_{n,1;i^\prime,i}\,\mathcal{G}^{(b_3,b_1)}_{n,2;j^\prime,j}\,\cdots,
    \end{split}\label{eq:tn}\\
    \begin{split}
    \mathcal{T}^{(b_1,b_2,b_3,b_4)}_{n;i,j,k,l}\equiv T^{(b_1,b_2,b_3,b_4)}_{i,j,k,l}
    \dd{\psi_{n,1}^{R(b_1-b_4)+p(i)}}\dd{\Bar{\psi}_{n,2}^{R(b_4-b_1)+p(i)}}\\
    \times\dd{\Bar{\chi}_{n,2}^{R(b_1-b_3)+p(j)}}\dd{\chi_{n,1}^{R(b_3-b_1)+p(j)}}
    \dd{\Bar{\psi}_{n,1}^{R(b_3-b_2)+p(k)}}\\\times\dd{\psi_{n,2}^{R(b_2-b_3)+p(k)}}
    \dd{\chi_{n,2}^{R(b_4-b_2)+p(l)}}\dd{\Bar{\chi}_{n,1}^{R(b_2-b_4)+p(l)}},
    \end{split}\label{eq:tsr}\\
    \begin{split}
    \mathcal{G}^{(b_1,b_4)}_{n,1;i^\prime,i}\equiv h_{i^\prime,i}^{(b_1,b_4)}\,
    \psi_{n,1}^{R(b_1-b_4)+p(i)}\Bar{\psi}_{n+\Hat{1},1}^{R(b_1-b_4)+p(i^\prime)}\\
    \times\psi_{n+\Hat{1},2}^{R(b_4-b_1)+p(i^\prime)}\Bar{\psi}_{n,2}^{R(b_4-b_1)+p(i)},
    \end{split}\label{}\\
    \begin{split}
    \mathcal{G}^{(b_3,b_1)}_{n,2;j^\prime,j}\equiv h_{j^\prime,j}^{(b_3,b_1)}\,
    \chi_{n,1}^{R(b_3-b_1)+p(j)}\Bar{\chi}_{n+\Hat{2},1}^{R(b_3-b_1)+p(j^\prime)}\\
    \times\chi_{n+\Hat{2},2}^{R(b_1-b_3)+p(j^\prime)}\Bar{\chi}_{n,2}^{R(b_1-b_3)+p(j)}.
    \end{split}\label{}
  \end{gather}
  $T^{(b_1,b_2,b_3,b_4)}$ is a decorated tensor kernel which includes no Grassmann number and its
  detail is given in Appendix~\ref{app:int}. $h_{i^\prime,i}^{(b_1,b_4)}$ is equal to the Kronecker delta
  $g_{i^\prime,i}$ for the initial tensor network. $R$ is a ramp function which satisfies
  \begin{equation}
    R(x)=
    \begin{cases}
      x,&x\geq 0\\
      0,&x<0
    \end{cases}.
  \end{equation}
  $p(i)$ equals to zero (one) when $i$ indicates a parity-even (-odd) state under a reflection.
  Note that $\{\psi_{n,\alpha}\}$, $\{\chi_{n,\alpha}\}$ and others are treated as independent variables.
  Figure~\ref{fig:dtn} illustrates Eq.~\eqref{eq:tn} as a network diagram.

  \subsection{Decorated TRG procedure}
  One TRG cycle consists of two steps, decomposing each tensor in Eq.~\eqref{eq:tn} by using the singular
  value decomposition (SVD) and contracting the decomposed tensors to obtain coarse-grained tensors.
  We extend an algorithm proposed in Ref.~\cite{Dittrich:2014mxa} to systems including fermions. Fermionic
  degrees of freedom are treated by following our previous approach given in Ref.~\cite{Shimizu:2014uva},
  but some modifications are introduced to preserve rotation and reflection symmetries.

  \begin{figure}
    \centering
    \includegraphics[width=75mm]{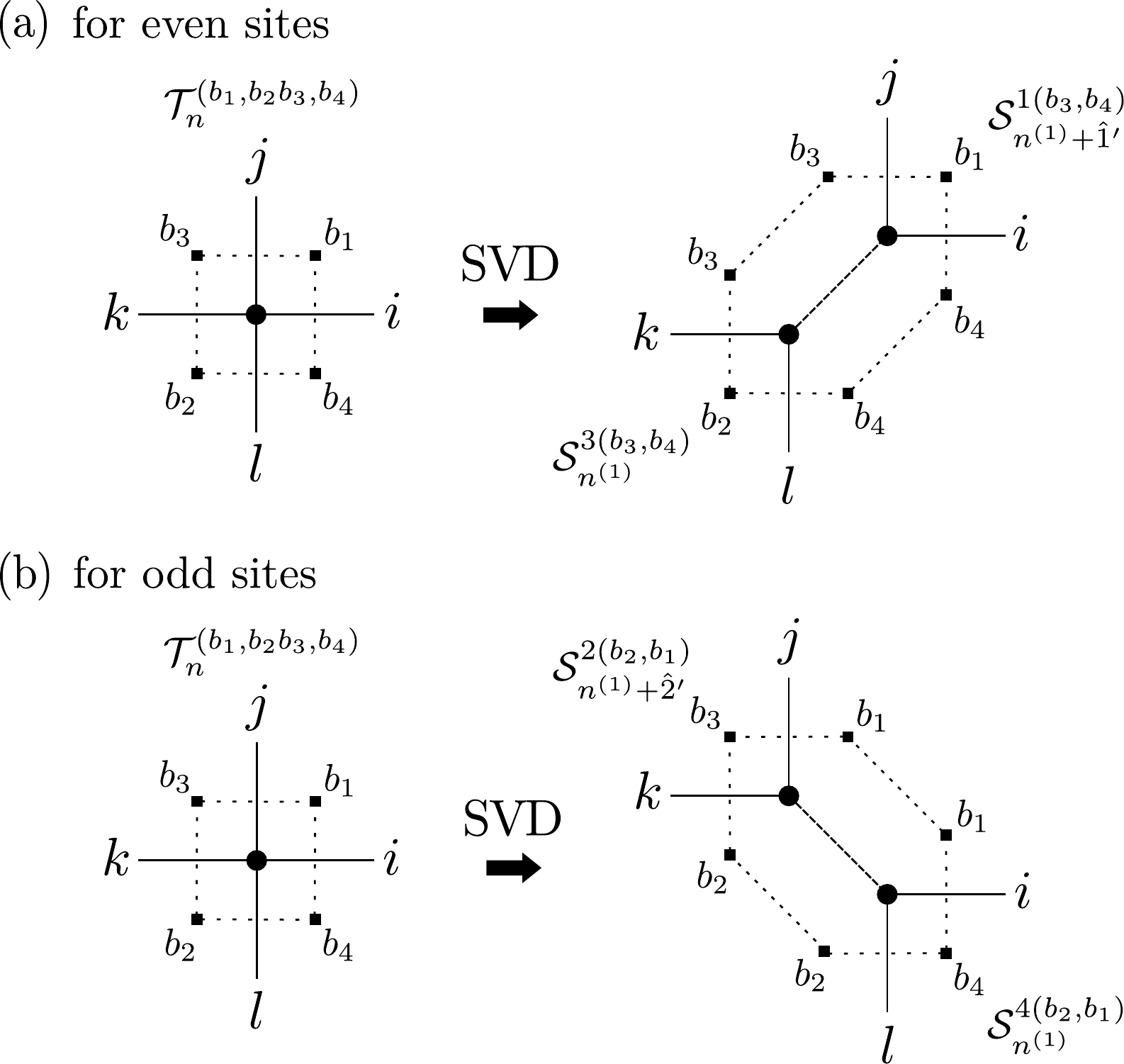}
    \caption{Schematic representation for the decompositions of (a) Eq.~\eqref{eq:gse} and (b) Eq.~\eqref{eq:gso}. Newly-introduced Grassmann variables
    $\left\{\Bar{\xi}_{n^{(1)}},\xi_{n^{(1)}},\Bar{\eta}_{n^{(1)}},\eta_{n^{(1)}}\right\}$ are assigned
    to the additional dashed lines.}
    \label{fig:svd}
  \end{figure}

  \begin{figure}
    \centering
    \includegraphics[width=70mm]{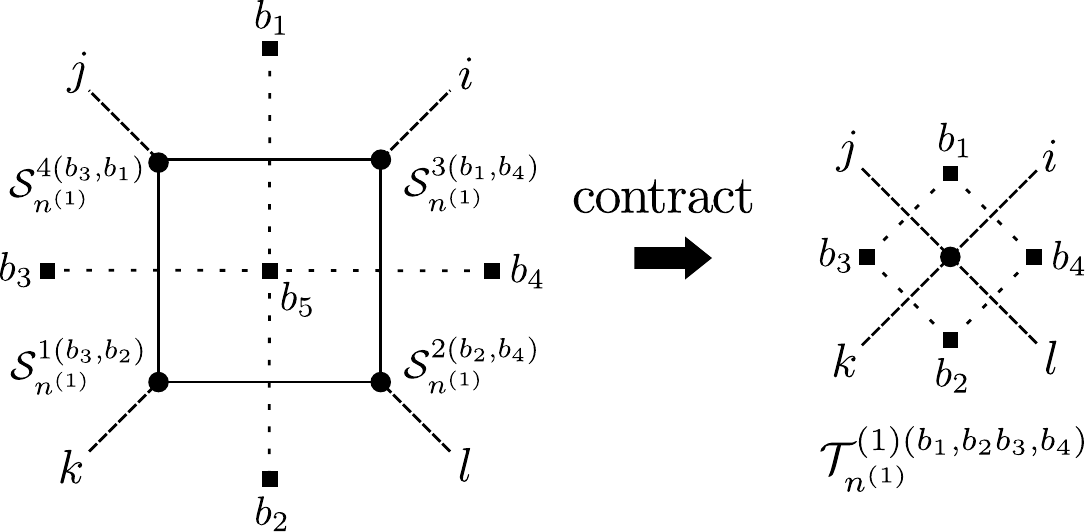}
    \caption{Schematic representation of the contraction of Eq.~\eqref{eq:ctr}. Solid lines in a closed loop,
    where old Grassmann variables $\left\{\Bar{\psi}_{n},\psi_{n},\Bar{\chi}_{n},\chi_{n}\right\}$ reside,
    can be contracted and replaced by $\mathcal{T}^{(1)}_{n^{(1)}}$.}
    \label{fig:ctr}
  \end{figure}

  First, let us explain the decomposition part. A matrix form of the decorated tensor kernel
  $T^{(b_1,b_2,b_3,b_4)}$ is defined as follows:
  \begin{equation}
    M^{(b_3,b_4)}_{(b_1;i,j),(b_2;k,l)}\equiv T^{(b_1,b_2,b_3,b_4)}_{i,j,k,l}.
    \label{}
  \end{equation}
  Then, we perform the SVD:
  \begin{equation}
    M^{(b_3,b_4)}_{(b_1;i,j),(b_2;k,l)}=\sum_m
    U^{(b_3,b_4)}_{(b_1;i,j),m}\,\sigma^{(b_3,b_4)}_{m}\,V^{(b_3,b_4)}_{m,(b_2;k,l)}.
    \label{eq:svd}
  \end{equation}
  Only the largest $D$ singular values out of $\left\{ \sigma^{(b_3,b_4)}_m \right\}$ are kept and others are
  approximated to zero.
  When $b_3=b_4$, $M^{(b_3,b_4)}$ is a symmetric matrix and the eigenvalue decomposition is applied rather
  than the SVD, that is, $V^{(b_3,b_4)}_{m,(b_2;k,l)}=U^{(b_3,b_4)}_{(b_2;k,l),m}$ and $\sigma^{(b_3,b_4)}_m$
  can be negative. The truncation is performed in that case, taking the absolute values of
  $\left\{\sigma^{(b_3,b_4)}_m \right\}$ into consideration. In the case $b_3=-b_4$, thanks to the reflection
  symmetry with respect to the $1^\prime$-axis (see Appendix~\ref{app:sym}), $M^{(b_3,b_4)}$ can be
  block-diagonalized. To derive a block-diagonalized form, we introduce special indices $O$, $A$, $B$,
  $O^\prime$, $A^\prime$ and $B^\prime$,
  \begin{align}
    O&\in \Set{(0;i,j)|i=j},\\
    A&\in \Set{(0;i,j)|i>j}\cup\Set{(b_1;i,j)|b_1>0},\\
    B&\in \Set{(0;i,j)|i<j}\cup\Set{(b_1;i,j)|b_1<0},\\
    O^\prime&\in \Set{(0;k,l)|k=l},\\
    A^\prime&\in \Set{(0;k,l)|k>l}\cup\Set{(b_2;k,l)|b_2>0},\\
    B^\prime&\in \Set{(0;k,l)|k<l}\cup\Set{(b_2;k,l)|b_2<0}.
    \label{}
  \end{align}
  Then, $M^{(b_3,b_4)}$ is block-diagonalized as follows:
  \begin{widetext}
  \begin{equation}
  \begin{split}
    &
    \begin{pmatrix}
      1&0&0\\
      0&\frac{1}{\sqrt{2}}&\frac{(-1)^{p(i)+p(j)+b_1}}{\sqrt{2}}\\
      0&\frac{1}{\sqrt{2}}&\frac{(-1)^{p(i)+p(j)+b_1+1}}{\sqrt{2}}
    \end{pmatrix}
    \begin{pmatrix}
      M_{OO^\prime}&M_{OA^\prime}&M_{OB^\prime}\\
      M_{AO^\prime}&M_{AA^\prime}&M_{AB^\prime}\\
      M_{BO^\prime}&M_{BA^\prime}&M_{BB^\prime}
    \end{pmatrix}
    \begin{pmatrix}
      1&0&0\\
      0&\frac{1}{\sqrt{2}}&\frac{1}{\sqrt{2}}\\
      0&\frac{(-1)^{p(k)+p(l)+b_2}}{\sqrt{2}}&\frac{(-1)^{p(k)+p(l)+b_2+1}}{\sqrt{2}}
    \end{pmatrix}\\
    =&
    \begin{pmatrix}
      M_{OO^\prime}&\sqrt{2}M_{OA^\prime}&0\\
      \sqrt{2}M_{AO^\prime}&M_{AA^\prime}+(-1)^{p(i)+p(j)+b_1}M_{BA^\prime}&0\\
      0&0&M_{AA^\prime}-(-1)^{p(i)+p(j)+b_1}M_{BA^\prime}
    \end{pmatrix},
    \label{}
  \end{split}
  \end{equation}
  \end{widetext}
  where the superscript $(b_3,b_4)$ is omitted. When $b_3$ is a odd (even) number, the upper-left
  (lower-right) block is parity-odd under the reflection. If the index $m$ which is introduced in the SVD,
  Eq.~\eqref{eq:svd} is in the parity-odd block, $p(m)=1$; otherwise $p(m)=0$. By using the above result, we
  define a decomposition of the Grassmann-valued tensor $\mathcal{T}^{(b_1,b_2,b_3,b_4)}_{n}$ on even sites,
  \begin{widetext}
  \begin{equation}
      \mathcal{T}^{(b_1,b_2,b_3,b_4)}_{n;i,j,k,l}=\sum_{m,m^\prime}\int
      \mathcal{S}^{1(b_3,b_4)}_{n^{(1)}+\Hat{1}^\prime;(b_1;i,j),m^\prime}\,
      \mathcal{S}^{3(b_3,b_4)}_{n^{(1)};(b_2;k,l),m}\,
      \mathcal{G}^{(1)(b_3,b_4)}_{n^{(1)},1^\prime;m^\prime,m},
    \label{eq:gse}
  \end{equation}
  with
  \begin{gather}
    \mathcal{G}^{(1)(b_3,b_4)}_{n^{(1)},1^\prime;m^\prime,m}\equiv
    h^{(1)(b_3,b_4)}_{m^\prime,m}\,\xi_{n^{(1)},1}^{R(b_3-b_4)+p(m)}
    \Bar{\xi}_{n^{(1)}+\Hat{1}^\prime,1}^{R(b_3-b_4)+p(m^\prime)}
    \xi_{n^{(1)}+\Hat{1}^\prime,2}^{R(b_4-b_3)+p(m^\prime)}\Bar{\xi}_{n^{(1)},2}^{R(b_4-b_3)+p(m)},\\
    h^{(1)(b_3,b_4)}_{m^\prime,m}\equiv\text{sgn}\left(\sigma^{(b_3,b_4)}_{m}\right)\delta_{m^\prime,m},\\
    \begin{split}
      \mathcal{S}^{1(b_3,b_4)}_{n^{(1)}+\Hat{1}^\prime;(b_1;i,j),m^\prime}\equiv
      S^{1(b_3,b_4)}_{(b_1;i,j),m^\prime}
      \dd{\Bar{\xi}_{n^{(1)}+\Hat{1}^\prime,1}^{R(b_3-b_4)+p(m^\prime)}}
      \dd{\xi_{n^{(1)}+\Hat{1}^\prime,2}^{R(b_4-b_3)+p(m^\prime)}}
      \dd{\psi_{n,1}^{R(b_1-b_4)+p(i)}}\dd{\Bar{\psi}_{n,2}^{R(b_4-b_1)+p(i)}}\\
      \dd{\Bar{\chi}_{n,2}^{R(b_1-b_3)+p(j)}}\dd{\chi_{n,1}^{R(b_3-b_1)+p(j)}},
    \end{split}\label{eq:s1}\\
    S^{1(b_3,b_4)}_{(b_1;i,j),m^\prime}\equiv
    (-1)^{R(b_1-b_3)+R(b_1-b_4)+R(b_3-b_4)}
    \sqrt{\left|\sigma^{(b_3,b_4)}_{m^\prime}\right|}\,V^{(b_4,b_3)}_{m^\prime,(b_1;i,j)},\\
    \begin{split}
      \mathcal{S}^{3(b_3,b_4)}_{n^{(1)};(b_2;k,l),m}\equiv
      S^{3(b_3,b_4)}_{(b_2;k,l),m}
      \dd{\xi_{n^{(1)},1}^{R(b_3-b_4)+p(m)}}
      \dd{\Bar{\xi}_{n^{(1)},2}^{R(b_4-b_3)+p(m)}}
      \dd{\Bar{\psi}_{n,1}^{R(b_3-b_2)+p(k)}}\dd{\psi_{n,2}^{R(b_2-b_3)+p(k)}}\\
      \dd{\chi_{n,2}^{R(b_4-b_2)+p(l)}}\dd{\Bar{\chi}_{n,1}^{R(b_2-b_4)+p(l)}},
    \end{split}\label{eq:s3}\\
    S^{3(b_3,b_4)}_{(b_2;k,l),m}\equiv
    \sqrt{\left|\sigma^{(b_3,b_4)}_{m}\right|}\,V^{(b_3,b_4)}_{m,(b_2;k,l)},
  \end{gather}
  \end{widetext}
  where $\left\{\Bar{\xi}_{n^{(1)}},\xi_{n^{(1)}}\right\}$ are newly-introduced Grassmann variables with
  $n^{(1)}$ a coarse-grained lattice site and $\text{sgn}$ is a sign function. This decomposition is schematically expressed in Fig.~\ref{fig:svd}~(a). It should be remarked that
  differently from our previous approach, $\xi_{n^{(1)}}$ is a two-component Grassmann variable and that is
  essential to preserve rotation and reflection symmetries. In Eq.~(\ref{eq:s1}), the $\pi$-rotation symmetry
  explained in Appendix~\ref{app:sym} is employed and it is assumed that $\xi_{n^{(1)}}$ transforms under
  rotations as $\psi_n$ does.  A decomposition for odd sites is defined as follows by exploiting the rotation
  symmetry discussed in Appendix~\ref{app:sym}:
  \begin{widetext}
  \begin{equation}
      \mathcal{T}^{(b_1,b_2,b_3,b_4)}_{n;i,j,k,l}=\sum_{m,m^\prime}\int
      \mathcal{S}^{2(b_2,b_1)}_{n^{(1)}+\Hat{2}^\prime;(b_3;j,k),m^\prime}\,
      \mathcal{S}^{4(b_2,b_1)}_{n^{(1)};(b_4;l,i),m}\,
      \mathcal{G}^{(1)(b_2,b_1)}_{n^{(1)},2^\prime;m^\prime,m},
    \label{eq:gso}
  \end{equation}
  with
  \begin{gather}
    \mathcal{G}^{(1)(b_2,b_1)}_{n^{(1)},2^\prime;m^\prime,m}\equiv
    h^{(1)(b_2,b_1)}_{m^\prime,m}\,\eta_{n^{(1)},1}^{R(b_2-b_1)+p(m)}
    \Bar{\eta}_{n^{(1)}+\Hat{2}^\prime,1}^{R(b_2-b_1)+p(m^\prime)}
    \eta_{n^{(1)}+\Hat{2}^\prime,2}^{R(b_1-b_2)+p(m^\prime)}\Bar{\eta}_{n^{(1)},2}^{R(b_1-b_2)+p(m)},\\
    \begin{split}
      \mathcal{S}^{2(b_2,b_1)}_{n^{(1)}+\Hat{2}^\prime;(b_3;j,k),m^\prime}\equiv
      S^{2(b_2,b_1)}_{(b_3;j,k),m^\prime}
      \dd{\eta_{n^{(1)}+\Hat{2}^\prime,2}^{R(b_1-b_2)+p(m^\prime)}}
      \dd{\Bar{\eta}_{n^{(1)}+\Hat{2}^\prime,1}^{R(b_2-b_1)+p(m^\prime)}}
      \dd{\Bar{\chi}_{n,2}^{R(b_1-b_3)+p(j)}}\dd{\chi_{n,1}^{R(b_3-b_1)+p(j)}}\\
      \dd{\Bar{\psi}_{n,1}^{R(b_3-b_2)+p(k)}}\dd{\psi_{n,2}^{R(b_2-b_3)+p(k)}},
    \end{split}\label{eq:s2}\\
     S^{2(b_2,b_1)}_{(b_3;j,k),m^\prime}\equiv (-1)^{R(b_2-b_3)}
     \sqrt{\left|\sigma^{(b_2,b_1)}_{m^\prime}\right|}\,V^{(b_1,b_2)}_{m^\prime,(b_3;j,k)},\\
    \begin{split}
      \mathcal{S}^{4(b_2,b_1)}_{n^{(1)};(b_4;l,i),m}\equiv
      S^{4(b_2,b_1)}_{(b_4;l,i),m}
      \dd{\Bar{\eta}_{n^{(1)},2}^{R(b_1-b_2)+p(m)}}
      \dd{\eta_{n^{(1)},1}^{R(b_2-b_1)+p(m)}}
      \dd{\chi_{n,2}^{R(b_4-b_2)+p(l)}}\dd{\Bar{\chi}_{n,1}^{R(b_2-b_4)+p(l)}}\\
      \dd{\psi_{n,1}^{R(b_1-b_4)+p(i)}}\dd{\Bar{\psi}_{n,2}^{R(b_4-b_1)+p(i)}},
    \end{split}\label{eq:s4}\\
    S^{4(b_2,b_1)}_{(b_4;l,i),m}\equiv (-1)^{R(b_4-b_2)+R(b_1-b_2)}
    \sqrt{\left|\sigma^{(b_2,b_1)}_{m}\right|}\,V^{(b_2,b_1)}_{m,(b_4;l,i)},
  \end{gather}
  \end{widetext}
  where $\left\{\Bar{\eta}_{n^{(1)}},\eta_{n^{(1)}}\right\}$ are also newly-introduced Grassmann variables
  and transform under rotations as $\chi_n$ does. This decomposition is schematically expressed in Fig.~\ref{fig:svd}~(b).

  Next, we perform a contraction for $\mathcal{S}^{1(b_3,b_2)}_{n^{(1)}}$,
  $\mathcal{S}^{2(b_3,b_1)}_{n^{(1)}}$, $\mathcal{S}^{3(b_1,b_4)}_{n^{(1)}}$ and
  $\mathcal{S}^{4(b_2,b_4)}_{n^{(1)}}$ as depicted in Fig.~\ref{fig:ctr}.
  The old Grassmann variables $\left\{\Bar{\psi}_n,\psi_n,\Bar{\xi}_n,\xi_n\right\}$ reside on a small closed
  loop in the Figure and we can integrate out all of them simultaneously with the contraction. As a result,
  we get a coarse-grained tensor
  \begin{widetext}
  \begin{equation}
  \begin{split}
    \mathcal{T}^{(1)(b_1,b_2,b_3,b_4)}_{n^{(1)};i,j,k,l}&=
    \sum_{b_5,p,q,r,s,p^\prime,q^\prime,r^\prime,s^\prime}\int
    \mathcal{S}^{1(b_3,b_2)}_{n^{(1)};(b_5;p,q),k}\,
    \mathcal{S}^{2(b_2,b_4)}_{n^{(1)};(b_5;s,p^\prime),l}\,
    \mathcal{S}^{3(b_1,b_4)}_{n^{(1)};(b_5;r^\prime,s^\prime),i}\,
    \mathcal{S}^{4(b_3,b_1)}_{n^{(1)};(b_5;q^\prime,r),j}\\
    &\quad\mathcal{G}^{(b_3,b_5)}_{n,2;q^\prime,q}\,
    \mathcal{G}^{(b_1,b_5)}_{n+\Hat{2},1;r^\prime,r}\,
    \mathcal{G}^{(b_5,b_4)}_{n+\Hat{1},2;s^\prime,s}\,
    \mathcal{G}^{(b_5,b_2)}_{n,1;p^\prime,p}\\
    &=\sum_{b_5,p,q,r,s,p^\prime,q^\prime,r^\prime,s^\prime}
    (-1)^{R(b_1-b_5)+R(b_2-b_5)+R(b_3-b_5)+R(b_4-b_5)+b_4+b_5}\\
    &\quad
    S^{1(b_3,b_2)}_{(b_5;p,q),k}\,
    S^{2(b_2,b_4)}_{(b_5;s,p^\prime),l}\,
    S^{3(b_1,b_4)}_{(b_5;r^\prime,s^\prime),i}\,
    S^{4(b_3,b_1)}_{(b_5;q^\prime,r),j}\,
    h^{(b_3,b_5)}_{q^\prime,q}\,
    h^{(b_1,b_5)}_{r^\prime,r}\,
    h^{(b_5,b_4)}_{s^\prime,s}\,
    h^{(b_5,b_2)}_{p^\prime,p}\\
    &\quad
    \dd{\xi_{n^{(1)},1}^{R(b_1-b_4)+p(i)}}\dd{\Bar{\xi}_{n^{(1)},2}^{R(b_4-b_1)+p(i)}}
    \dd{\Bar{\eta}_{n^{(1)},2}^{R(b_1-b_3)+p(j)}}\dd{\eta_{n^{(1)},1}^{R(b_3-b_1)+p(j)}}\\
    &\quad
    \dd{\Bar{\xi}_{n^{(1)},1}^{R(b_3-b_2)+p(k)}}\dd{\xi_{n^{(1)},2}^{R(b_2-b_3)+p(k)}}
    \dd{\eta_{n^{(1)},2}^{R(b_4-b_2)+p(l)}}\dd{\Bar{\eta}_{n^{(1)},1}^{R(b_2-b_4)+p(l)}}\\
    &\equiv T^{(1)(b_1,b_2,b_3,b_4)}_{i,j,k,l}
    \dd{\xi_{n^{(1)},1}^{R(b_1-b_4)+p(i)}}\dd{\Bar{\xi}_{n^{(1)},2}^{R(b_4-b_1)+p(i)}}
    \dd{\Bar{\eta}_{n^{(1)},2}^{R(b_1-b_3)+p(j)}}\dd{\eta_{n^{(1)},1}^{R(b_3-b_1)+p(j)}}\\
    &\quad
    \dd{\Bar{\xi}_{n^{(1)},1}^{R(b_3-b_2)+p(k)}}\dd{\xi_{n^{(1)},2}^{R(b_2-b_3)+p(k)}}
    \dd{\eta_{n^{(1)},2}^{R(b_4-b_2)+p(l)}}\dd{\Bar{\eta}_{n^{(1)},1}^{R(b_2-b_4)+p(l)}},
  \end{split}\label{eq:ctr}
  \end{equation}
  \end{widetext}
  where only the new Grassmann variables remain.
  It is easily confirmed that the constraints on the tensor kernel $T^{(1)(b_1,b_2,b_3,b_4)}$ from the
  rotation symmetry are the same as those on the initial tensor given in Appendix~\ref{app:sym}. The
  constraints from the reflection symmetry are also derived by assuming appropriate transformality of
  $\xi_{n^{(1)}}$ and $\eta_{n^{(1)}}$. Roles of the $1$-axis ($2$-axis) and $2^\prime$-axis
  ($1^\prime$-axis) are swapped, compared with the case of the initial tensor.

  Further iterations can be performed in a similar manner. After $N\equiv \log_2L^2$ iterations with $L$ the
  lattice length, one $N$-times coarse-grained tensor $\mathcal{T}^{(N)(b_1,b_2,b_3,b_4)}_{n^{(N)}}$ covers
  whole lattice. Therefore, the density matrix $\rho$ is represented as
  \begin{equation}
  \begin{split}
    \rho_{(b_1;i),(b_2;k)}=&\sum_{j,j^\prime,k^\prime}\int
    \mathcal{T}^{(N)(b_1,b_2,b_2,b_1)}_{0;i,j,k^\prime,j^\prime}\\
    &\mathcal{G}^{(N)(b_2,b_2)}_{0,1;k^\prime,k}
    \mathcal{G}^{(N)(b_2,b_1)}_{0,2;j^\prime,j}\\
    =&\sum_{j}
    (-1)^{R(b_2-b_1)}\,
    T^{(N)(b_1,b_2,b_2,b_1)}_{i,j,k,j}\\
    &\text{sgn}\left(\sigma^{(N-1)(b_2,b_2)}_{k}\right)\text{sgn}\left(\sigma^{(N-1)(b_2,b_1)}_{j}\right),
  \end{split}\label{eq:dm}
  \end{equation}
  and the partition function is $Z=\tr\rho$
  where periodic boundary conditions are employed. The density matrix $\rho$ can be block-diagonalized
  because of the reflection symmetry with respect to the $1$-axis, namely the parity symmetry. That enables
  us to compute an expectation value of a projection operator into the parity-odd subspace as follows:
  \begin{equation}
  \begin{split}
    \langle P_\mathrm{odd}\rangle&=\frac{\tr(P_\mathrm{odd}\,\rho)}{Z}\\
    &=\frac{1}{Z}\Biggl\{\Biggl(\sum_{i:p(i)=1}\rho_{(0;i),(0;i)}\Biggr)\\
    &\quad +\Biggl[\sum_{b_1>0,i}\left(\rho_{(b_1;i),(b_1;i)}-(-1)^{p(i)}\rho_{(b_1;i),(-b_1;i)}\right)
    \Biggr]\Biggr\}.
  \end{split}\label{eq:po}
  \end{equation}
  Taking the large volume limit, it will be one half in the parity-broken phase; otherwise zero.

  \section{Numerical study}
  \label{sec:num}

  \subsection{Strong coupling limit}
\begin{figure}
    \centering
    \includegraphics[width=80mm]{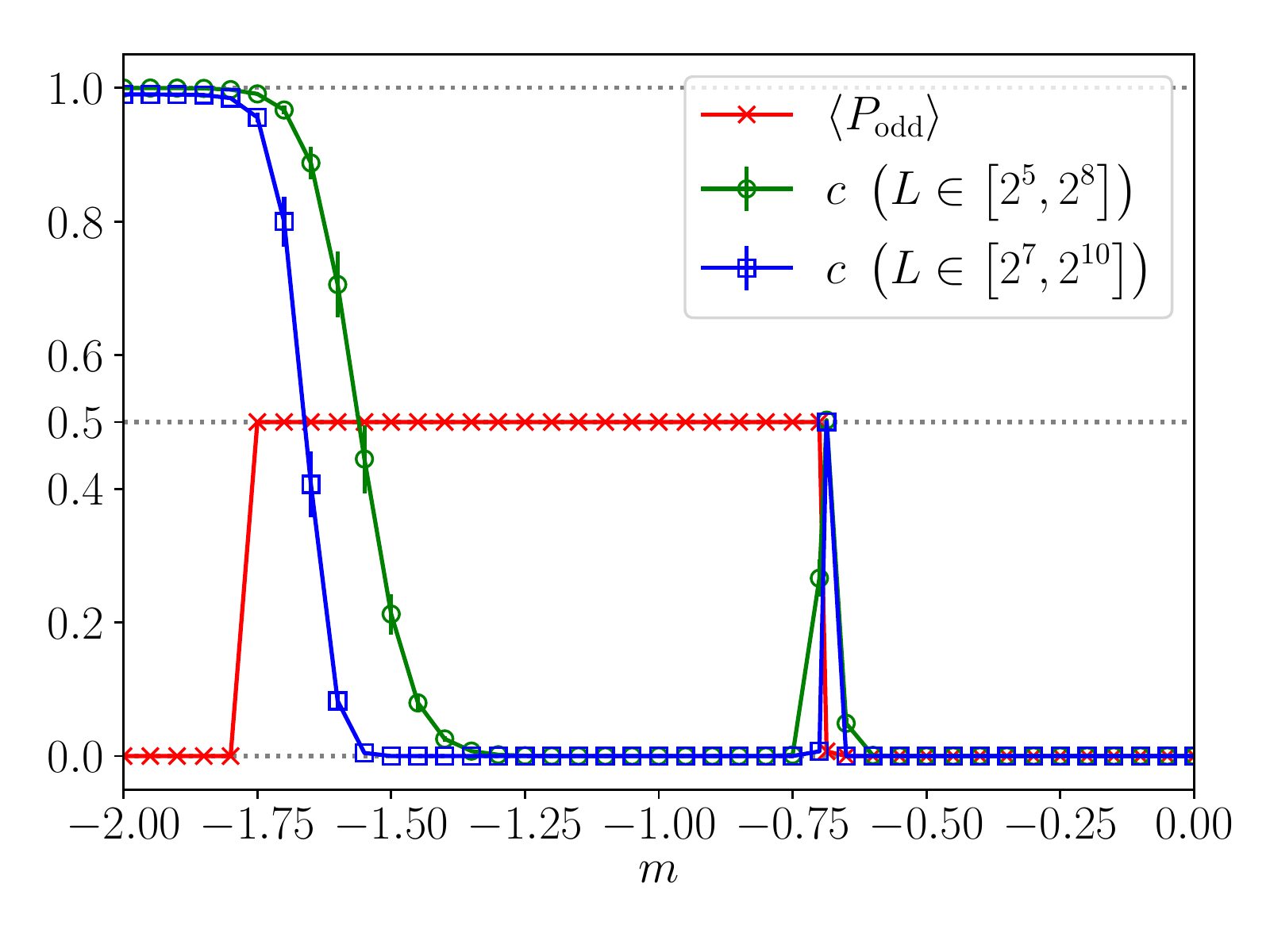}
    \caption{$\langle P_\mathrm{odd}\rangle$ and central charge $c$ as a function of fermion mass $m$ at
    $g=\infty$ evaluated with $D=160$. $\langle P_\mathrm{odd}\rangle$ is evaluated on a
    $L=2^{20}$ lattice and $c$ is from $\chi^2$-fits with Eq.~\eqref{eq:l0} for $L\in \left[2^{5},
    2^{8}\right]$ (circle) or $L\in \left[2^{7},2^{10}\right]$ (square). Dotted lines denote $\langle P_\mathrm{odd}\rangle =0,0.5$ and $c=0,1$ to guide eyes.}
    \label{fig:pocc_scl}
  \end{figure}

\begin{figure}
    \centering
    \includegraphics[width=80mm]{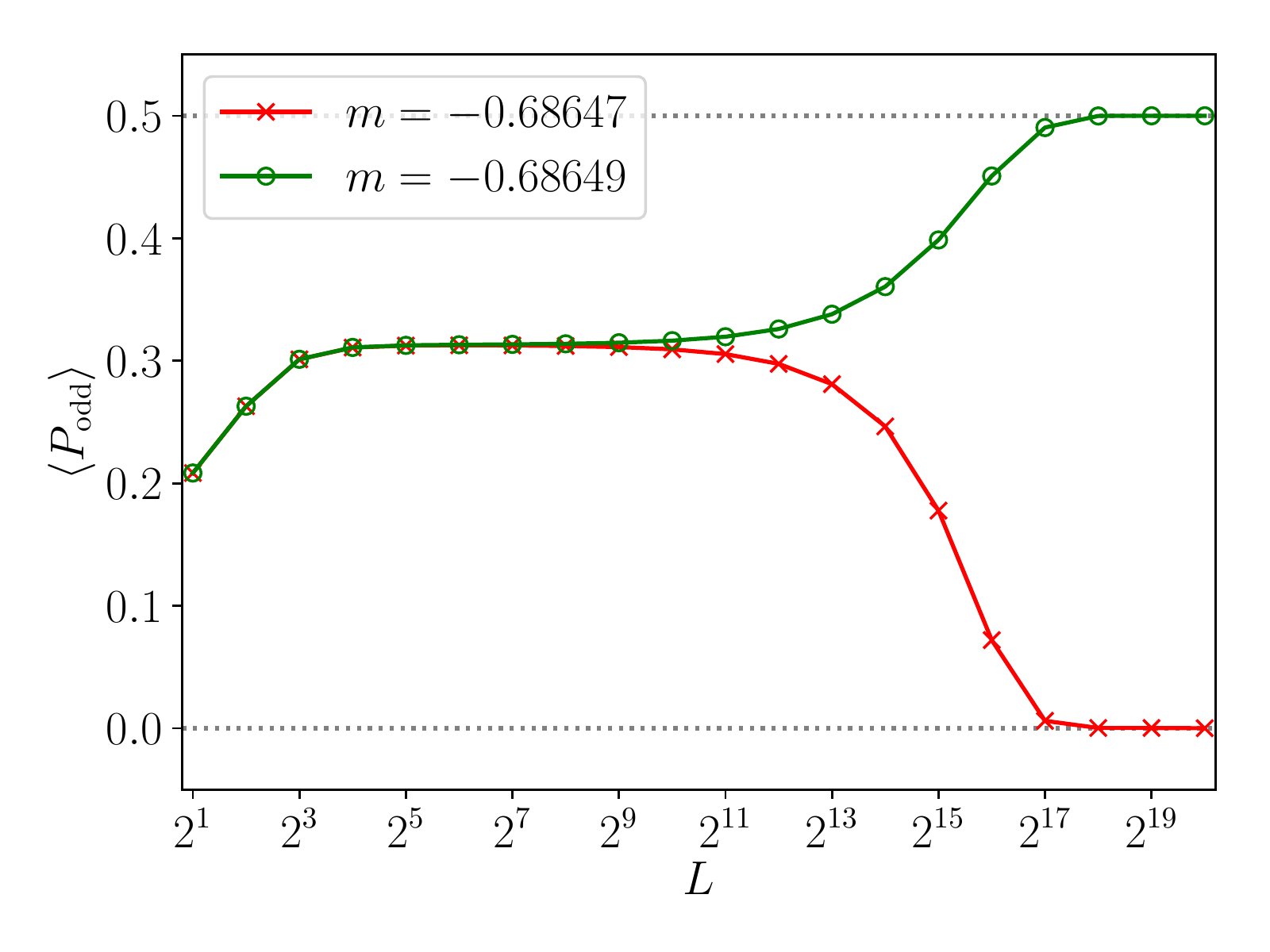}
    \caption{Convergence of $\langle P_\mathrm{odd}\rangle$ at $g=\infty$ evaluated with $D=160$ as a
    function of $L$. The result of $m=-0.68647$ goes to zero (symmetric phase) while that of $m=-0.68649$
    reaches one half (parity-broken phase). Dotted lines denote $\langle P_\mathrm{odd}\rangle =0,0.5$ to guide eyes.}
    \label{fig:fl_scl}
  \end{figure}

\begin{figure}
    \centering
    \includegraphics[width=80mm]{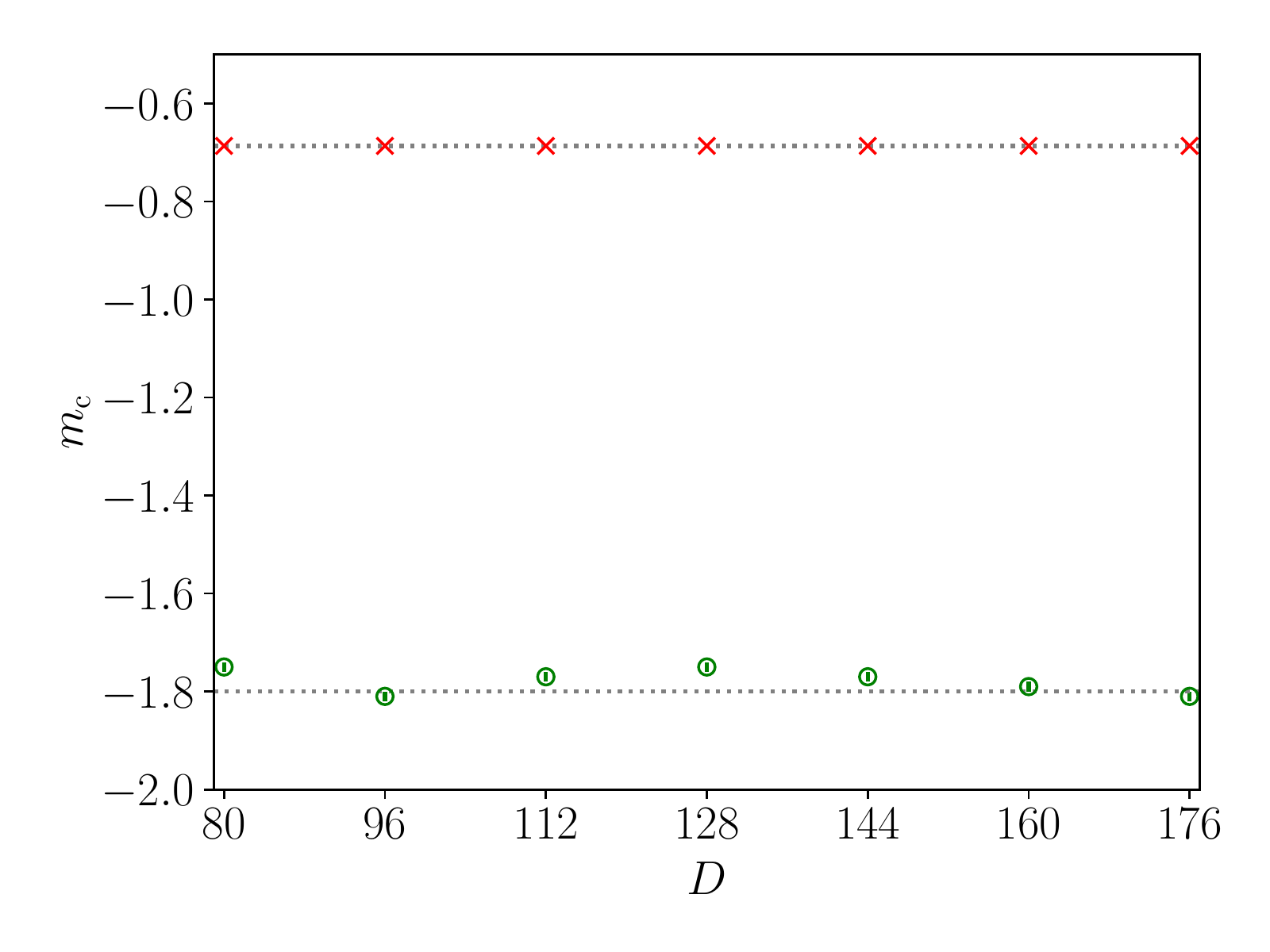}
    \caption{Convergence of two phase transition points evaluated by a $\langle P_\mathrm{odd}\rangle$
    analysis on a $L=2^{20}$ lattice at $g=\infty$ as a function of the SVD truncation number $D$. Dotted lines are for guiding eyes.}
    \label{fig:tp_scl}
  \end{figure}

  Before investigation at finite couplings, we first examine the strong coupling limit. In this case,
  the gauge action of Eq.~\eqref{eq:ga} is dropped and the conventional TRG is applied rather than the decorated
  TRG as in Ref.~\cite{Shimizu:2014uva}. Taking advantage of the symmetry-preserving
  algorithm, we calculate $\langle P_\mathrm{odd}\rangle$ according to Eq.~\eqref{eq:po} in order to detect a
  parity-broken phase in an explicit manner, while we employed the analyses of the Lee-Yang/Fisher zeros and chiral susceptibility in Ref.~\cite{Shimizu:2014uva}.
  Figure~\ref{fig:pocc_scl} shows our estimate of $\langle P_\mathrm{odd}\rangle$ on a $L=2^{20}$ lattice with
  $D=160$ as the SVD truncation parameter. It overtly indicates a parity-breaking transition around $m=-0.7$. We present the convergence behavior of $\langle P_\mathrm{odd}\rangle$ with
  increasing $L$ around the transition point in Fig.~\ref{fig:fl_scl}. The $D$ dependence of the transition point is given in
  Fig.~\ref{fig:tp_scl}, where good convergence is observed for $D\geq 80$ within the discretized resolution of $\Delta m=0.00001$.
  We conclude that the critical mass is $m_\mathrm{c}=-0.68648(1)$, which is consistent with values reported in
  Ref.~\cite{Gausterer:1995np,Wenger:2008tq} and also our previous result~\cite{Shimizu:2014uva}.

  \begin{figure}
    \centering
    \includegraphics[width=80mm]{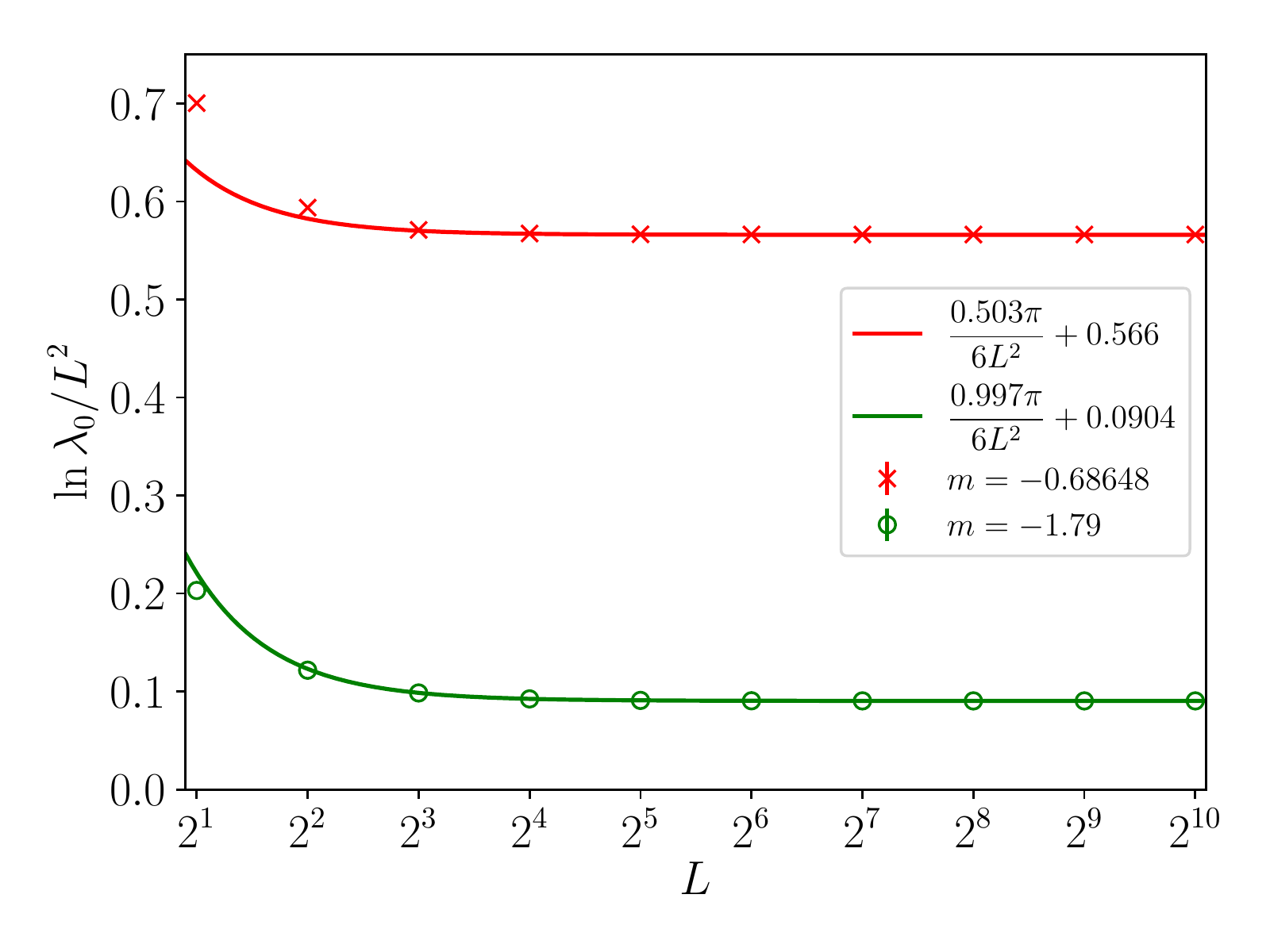}
    \caption{The largest eigenvalue $\lambda_0$ of the density matrix $\rho$ at $g=\infty$
    evaluated with $D=160$ as a function of $L$. Two cases of $m=-0.68648$ and $m=-1.79$ are presented.
    Solid curves represent fit results with Eq.~\eqref{eq:l0} for $L\in\left[2^5,2^8\right]$.}
    \label{fig:cc_scl}
  \end{figure}

  \begin{table}
    \centering
    \caption{Fit results of $\lambda_0$ with Eq.~\eqref{eq:l0} at $g=\infty$ for $L\in\left[2^5,2^8\right]$.
    $\lambda_0$ is evaluated with $D=160$.}
    \label{tab:cc_scl}
    \begin{ruledtabular}
    \begin{tabular}{cccc}\toprule
      $m$&$c$&$f_\infty$&$\sqrt{\chi^2/\mathrm{d.o.f}}$\\ \hline
      $-0.68648$ &$0.50272(30)$&$-0.566320061(79)$&$4.5$ \\
      $-1.79$    &$0.99668(99)$&$-0.09038590(23)$ &$0.035$ \\ \bottomrule
    \end{tabular}
    \end{ruledtabular}
  \end{table}

  An important new finding is another phase transition around $m=-1.8$. It should be noted that this transition was
  not detected in our previous study using the chiral susceptibility analysis. One of the possible candidates is the
  Berezinskii-Kosterlitz-Thouless (BKT) transition which does not have singularity in the free energy.
  To clarify the properties of the transition, we measure the central charge $c$, which should be one at the BKT criticality,
  one half at the Ising criticality or zero at non-criticality. As demonstrated by Gu and Wen using the TRG in
  Ref.~\cite{Gu:2009aa}, one can evaluate it from the largest eigenvalue of the density matrix
  $\rho$ with the aid of CFT. Since the eigenvalues $\{\lambda_i\}$ are of a form
  \begin{equation}
    \lambda_i=e^{-f_\infty L^2-2\pi\left(\Delta_i-\frac{c}{12}\right)},
    \label{eq:li}
  \end{equation}
  where $f_\infty$ is a non-universal constant and $\{\Delta_i\}$ are scaling dimensions, we use the
  following equation for $\chi^2$-fit:
  \begin{equation}
    \ln\lambda_0=\frac{\pi c}{6}-f_\infty L^2.
    \label{eq:l0}
  \end{equation}
  Note that our algorithm given in Sec.~\ref{sec:met} does not preserve hermiticity of the density matrix
  and it is numerically broken. Because of this, we instead compute singular values as our estimate of
  eigenvalues. Figure~\ref{fig:cc_scl} shows $L$ dependence of the largest
  eigenvalue $\lambda_0$ at $m=-1.79$, $-0.68648$, whose errors are estimated by the difference between the results with $D=160$ and $80$, and Table~\ref{tab:cc_scl} summarizes the fit results, which indicate $c=1$ for $m=-1.79$ and $c=\frac{1}{2}$ for $m=-0.68648$. The $m$ dependence of the central charge is plotted in Fig.~\ref{fig:pocc_scl}.  We observe a wide range of $c=1$ for $m\lesssim-1.8$, while a
  clear $c=\frac{1}{2}$ peak is found around $m=-0.7$ of the Ising transition point. The existence of the critical region (not point) for $m\lesssim-1.8$ is consistent with an expected property of the BKT transition.  Note that the central charge in Fig.~\ref{fig:pocc_scl} rises up
  more gently than $\langle P_\mathrm{odd}\rangle$. One of the main reasons is that the central charge is estimated
  with data sets of somewhat smaller lattices in order to avoid accumulation of the SVD truncation error which makes the TRG
  flows go away from the critical surface.  The finite size effect is clearly observed by comparing two fit
  results between $L\in\left[2^5,2^8\right]$ and $L\in\left[2^7,2^{10}\right]$.
  On the other hand, $\langle P_\mathrm{odd}\rangle$ is a quantity related to a rather trivial fixed point
  which renormalization group flows finally reach away from criticality. It is less contaminated by
  the SVD truncation error so that we are allowed to take large $L$ limit more reliably. However,it should be noted that
  $\langle P_\mathrm{odd}\rangle$ is much more unstable around the BKT transition point than the Ising
  transition point: Our estimate of the BKT transition point by the
  $\langle P_\mathrm{odd}\rangle$ analysis has larger $D$ dependence as observed in Fig.~\ref{fig:tp_scl}.

  \begin{figure}
    \centering
    \includegraphics[width=80mm]{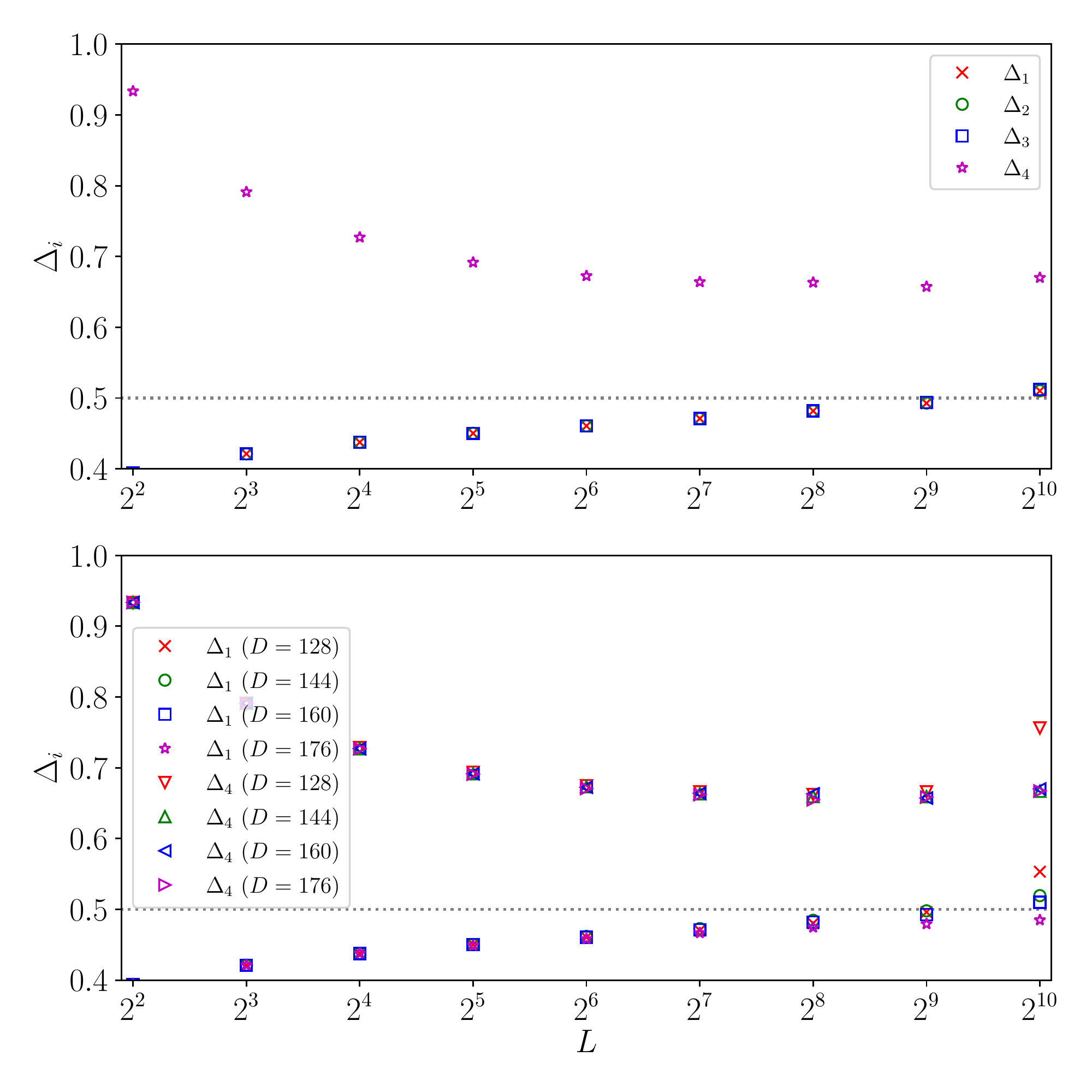}
    \caption{\textbf{Top:} scaling dimensions $\Delta_1$, $\Delta_2$, $\Delta_3$ and $\Delta_4$ at
    $(m,g)=(-2,\infty)$ as a function of $L$ evaluated with $D=160$. Dotted Line denotes $\Delta_i=0.5$ to guide eyes.
\textbf{Bottom:} $D$ dependence of $\Delta_1$ and $\Delta_4$. Dotted line denotes $\Delta_i=0.5$ to guide eyes.}
    \label{fig:sd_scl_bkt}
  \end{figure}

  \begin{figure}
    \centering
    \includegraphics[width=80mm]{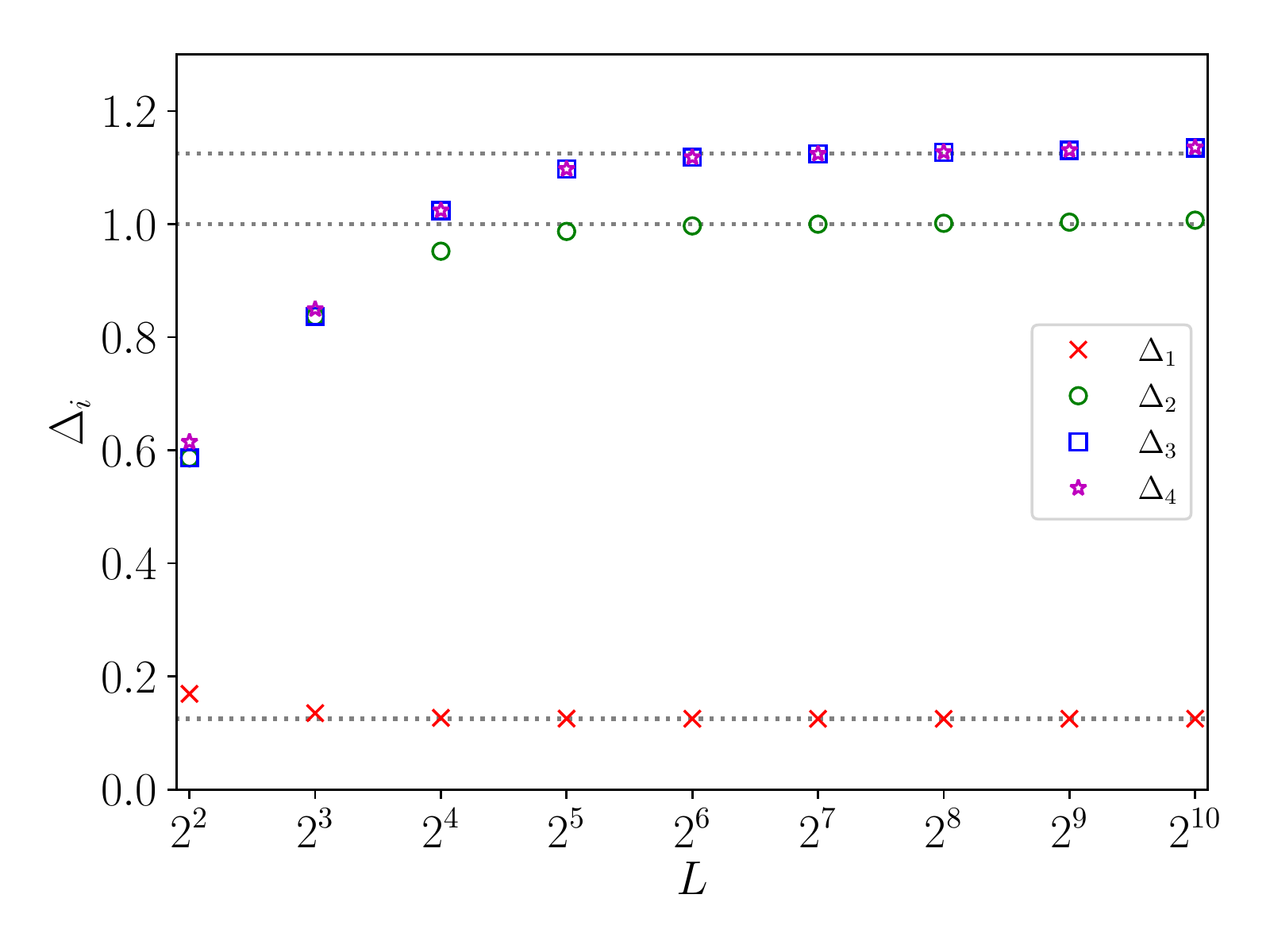}
    \caption{Scaling dimensions $\Delta_1$, $\Delta_2$, $\Delta_3$ and $\Delta_4$ at
    $(m,g)=(-0.68648,\infty)$ as a function of $L$ evaluated with $D=160$. Dotted lines denote $\Delta_1=\frac{1}{8}$, $\Delta_2=1$ and
  $\Delta_3=\Delta_4=\frac{9}{8}$ to guide eyes.}
    \label{fig:sd_scl_is}
  \end{figure}

  \begin{figure}
    \centering
    \includegraphics[width=80mm]{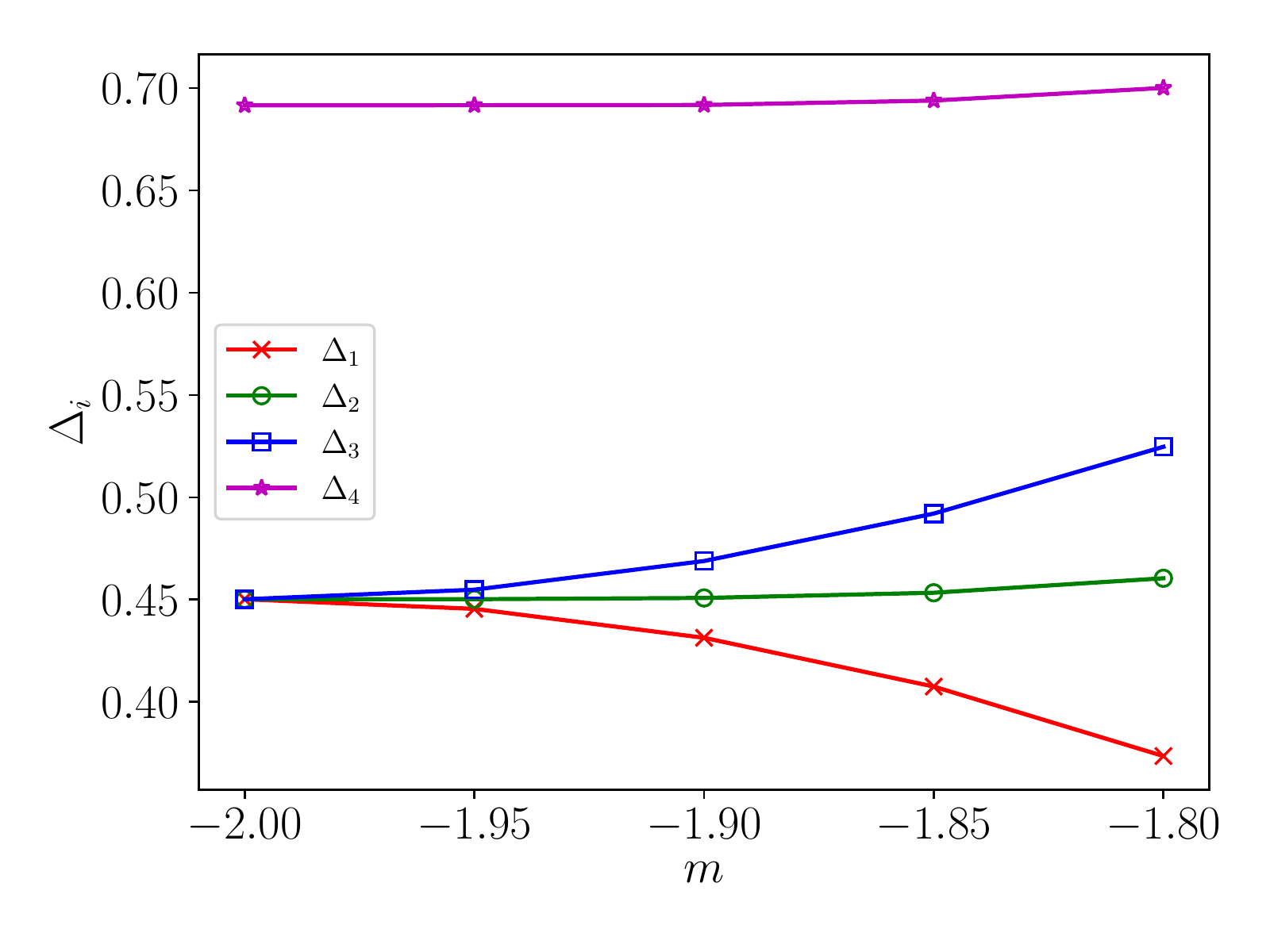}
    \caption{Scaling dimensions $\Delta_1$, $\Delta_2$, $\Delta_3$ and $\Delta_4$ at
    $g=\infty$ evaluated with $D=160$ on a $L=2^5$ lattice as a function of $m$ in the BKT phase.}
    \label{fig:msd_scl}
  \end{figure}

  We also measure scaling dimensions using Eq.~\eqref{eq:li}:
  \begin{equation}
    \Delta_i=\frac{1}{2\pi}\ln\left(\frac{\lambda_0}{\lambda_i}\right).
    \label{eq:di}
  \end{equation}
  In a BKT phase, $\{\Delta_i\}$ vary continuously due to the existence of an exactly marginal
  operator. At $m=-2$ the model is mapped to a critical six-vertex model which belongs to an universality
  class described by a free boson compactified on a circle with radius $R=\frac{1}{\sqrt{2}}$
  (see, e.g., Ref.~\cite{Ginsparg:1988ui}).
  In this case, the scaling dimensions are known to be
  \begin{equation}
    \Delta_i=\frac{M^2}{4R^2}+R^2N^2,\quad M,N\in\mathbb{Z},
    \label{}
  \end{equation}
  which means that all the lowest four nonzero scaling dimensions $\Delta_1$, $\Delta_2$, $\Delta_3$ and
  $\Delta_4$ are equal to one half. The top panel of Fig.~\ref{fig:sd_scl_bkt} shows $L$ dependence of the scaling dimensions. 
We find sizable deviation from one half, especially for $\Delta_4$, compared to the case of $(m,g)=(-0.68648,\infty)$
  in Fig.~\ref{fig:sd_scl_is}, where $\Delta_1=\frac{1}{8}$, $\Delta_2=1$ and
  $\Delta_3=\Delta_4=\frac{9}{8}$ is expected for the Ising transition. 
This is due to the well-known fact that the existence of an exactly marginal operator in the
  BKT transition causes an universal $\order{(\ln L)^{-1}}$ correction to the scaling dimensions while the central
  charge is affected by a smaller correction of $\order{(\ln L)^{-3}}$~\cite{Cardy:1986aa,
  Cardy:1987aa}. Although precise determination of the scaling dimensions requires much larger lattice size, accumulation of
  the SVD truncation error makes it difficult (see $D$ dependence of the scaling dimensions evaluated at $L=2^{10}$ in
the bottom panel of Fig.~\ref{fig:sd_scl_bkt}). 

Putting aside the
  logarithmic finite size corrections on the scaling dimensions, it should be important to point out that $\Delta_1$, $\Delta_2$ and $\Delta_3$
  form a triplet.  The free boson compactified with $R=\frac{1}{\sqrt{2}}$ corresponds to a CFT with
an  $\mathrm{SU}(2)$ symmetry and the triplet spectrum is ascribed to it. Note that the
  $S=\frac{1}{2}$ Heisenberg XXX model belongs to the same universality class and its scaling dimensions obtained
  by numerically solving the Bethe ansatz equations show similar behavior to our result~\cite{Avdeev:1990wh,Nomura:1993aa}: triplet spectrum close to the value of one half and singlet one deviated from the value of one half.
 A remaining question of interest is whether or not the
  $\mathrm{SU}(2)$ symmetry arises at $m=-2$ only. We present $m$ dependence of the scaling
dimensions in the BKT phase around $m=-2$ in Fig.~\ref{fig:msd_scl}, where we find that the triplet spectrum is observed 
only at $m=-2$.

  \subsection{Finite couplings}
\begin{figure}
    \centering
    \includegraphics[width=80mm]{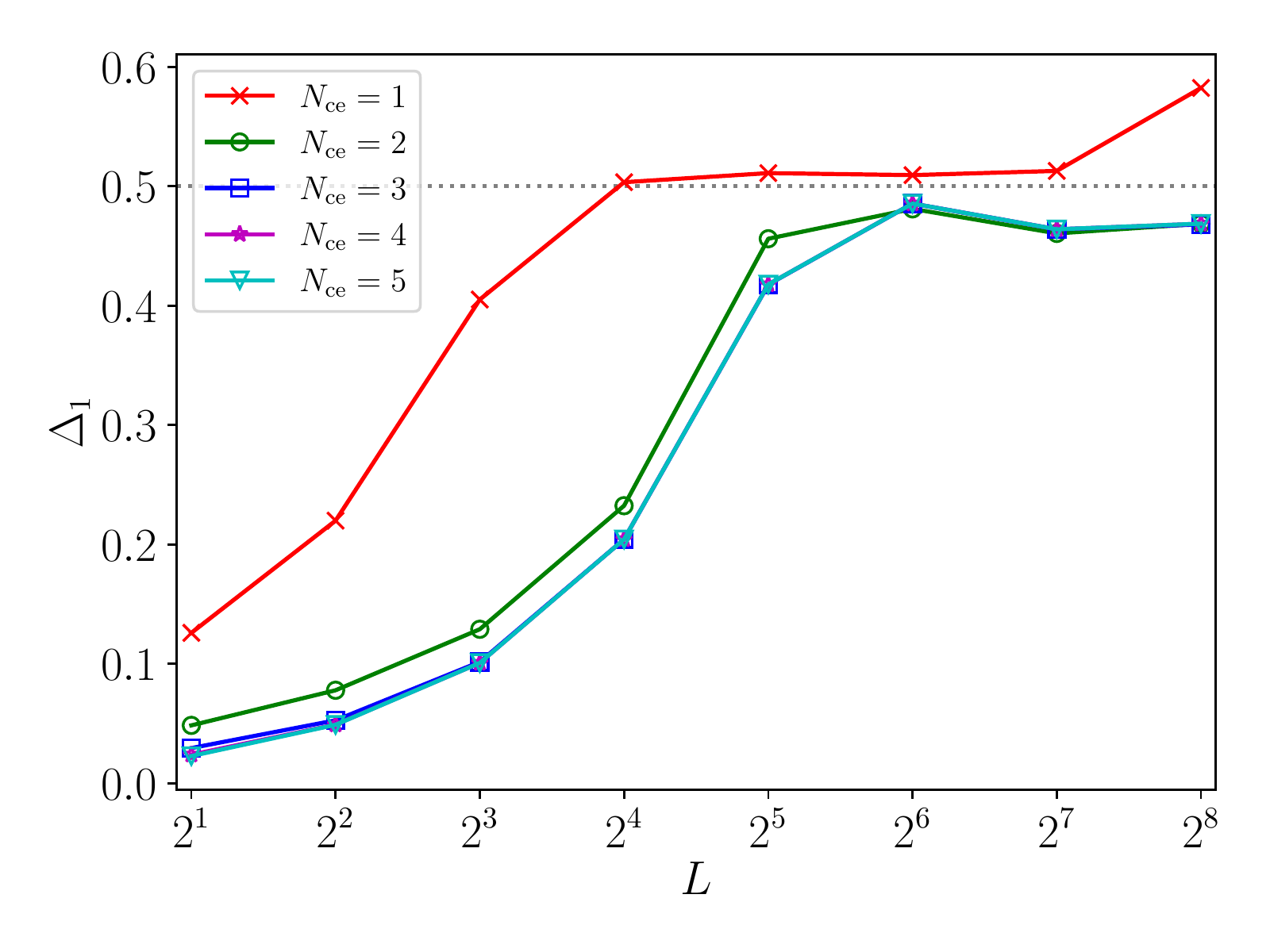}
    \caption{The lowest nonzero scaling dimension $\Delta_1$ evaluated at $(m,g)=(-2,0.1)$ with $D=48$ and $N_\mathrm{ce}=1,\cdots,5$ as a function of $L$. Dotted line denotes $\Delta_1=0.5$ to guide eyes.}
    \label{fig:ndep}
  \end{figure}

  We employ three couplings $g=1.0,\,0.5,\,0.1$ to investigate the phase structure
  in the finite-coupling region, where an additional parameter $N_\mathrm{ce}$ is required to control the truncation of the character expansion.
  In our previous approach with the conventional TRG method in Ref.~\cite{Shimizu:2014uva,Shimizu:2014fsa}, only the initial tensor size is related to the truncation number $N_\mathrm{ce}$ and we chose a somewhat excessive value to safely neglect the truncation error. 
However, the computational cost
  of the decorated TRG method heavily depends on $N_\mathrm{ce}$ and smaller values are preferred.
  Figure~\ref{fig:ndep} shows our estimate of $\Delta_1$ at $(m,g)=(-2,0.1)$ with several values of $N_\mathrm{ce}$. The result looks converged for $N_\mathrm{ce}\geq 3$ even on smaller lattices so that we employ
  $N_\mathrm{ce}=3$ in this work. Since the convergency property of the character expansion becomes better toward the strong coupling limit, the choice of $N_\mathrm{ce}\geq 3$ is also valid for the cases of $g=1.0,\,0.5$. It should be noted that our choice
  of the SVD truncation number $D$ in the finite coupling study is much smaller than in the case of the
  strong coupling limit with the conventional TRG method. In the case of the decorated TRG method, however, we eventually take $D\times(2N_\mathrm{ce}+1)^2$ singular values overall with the choice of $D$ singular values in each block.

\begin{figure}
    \centering
    \includegraphics[width=80mm]{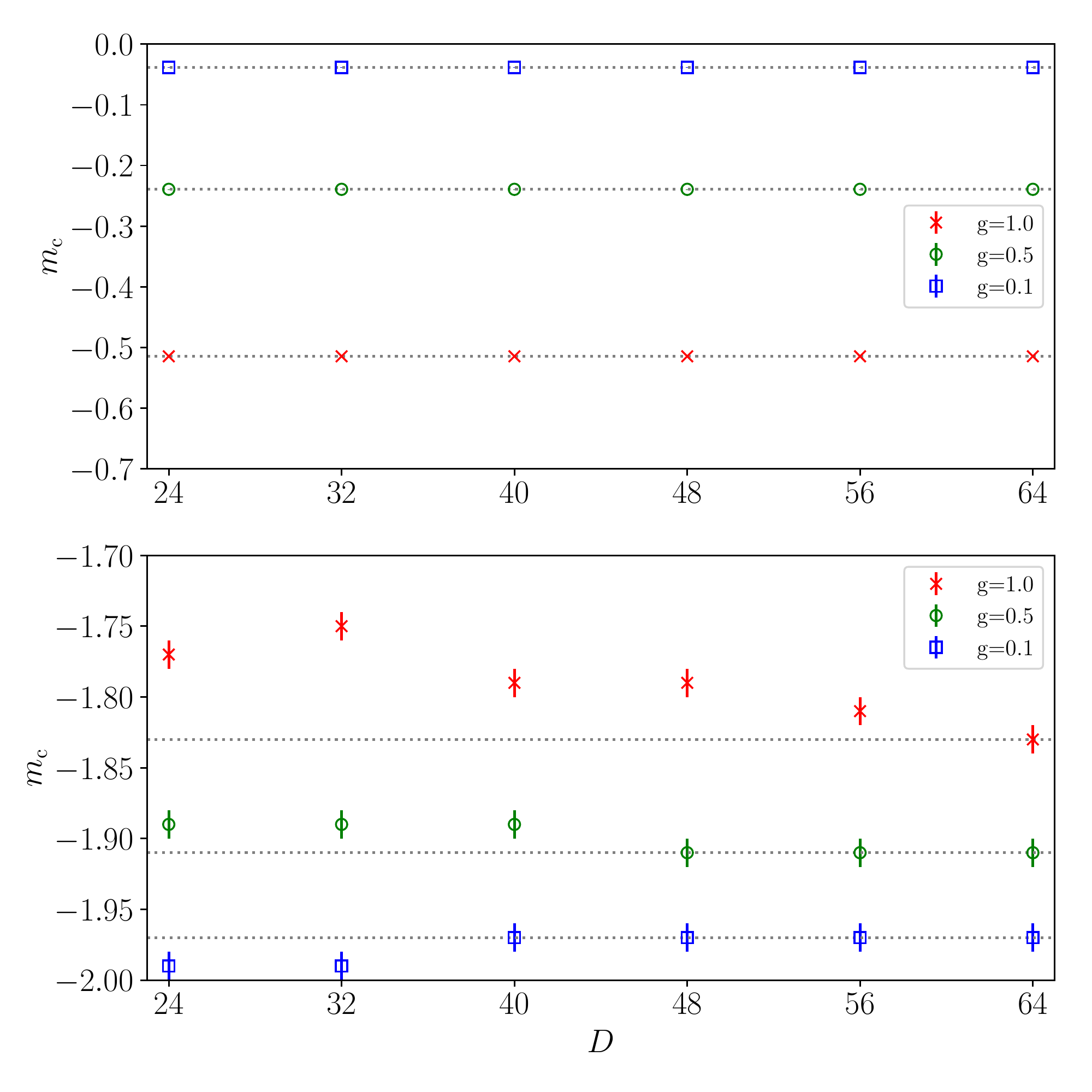}
    \caption{Convergence behavior of the Ising transition point (top) and the BKT transition point (bottom) as a function of the SVD truncation number $D$. $\langle P_\mathrm{odd}\rangle$ analysis is made at $g=1.0,\,0.5,\,0.1$ on a $L=2^{20}$ lattice with $N_\mathrm{ce}=3$. Dotted lines are for guiding eyes.}
    \label{fig:tp_fc}
  \end{figure}

\begin{figure}
    \centering
    \includegraphics[width=80mm]{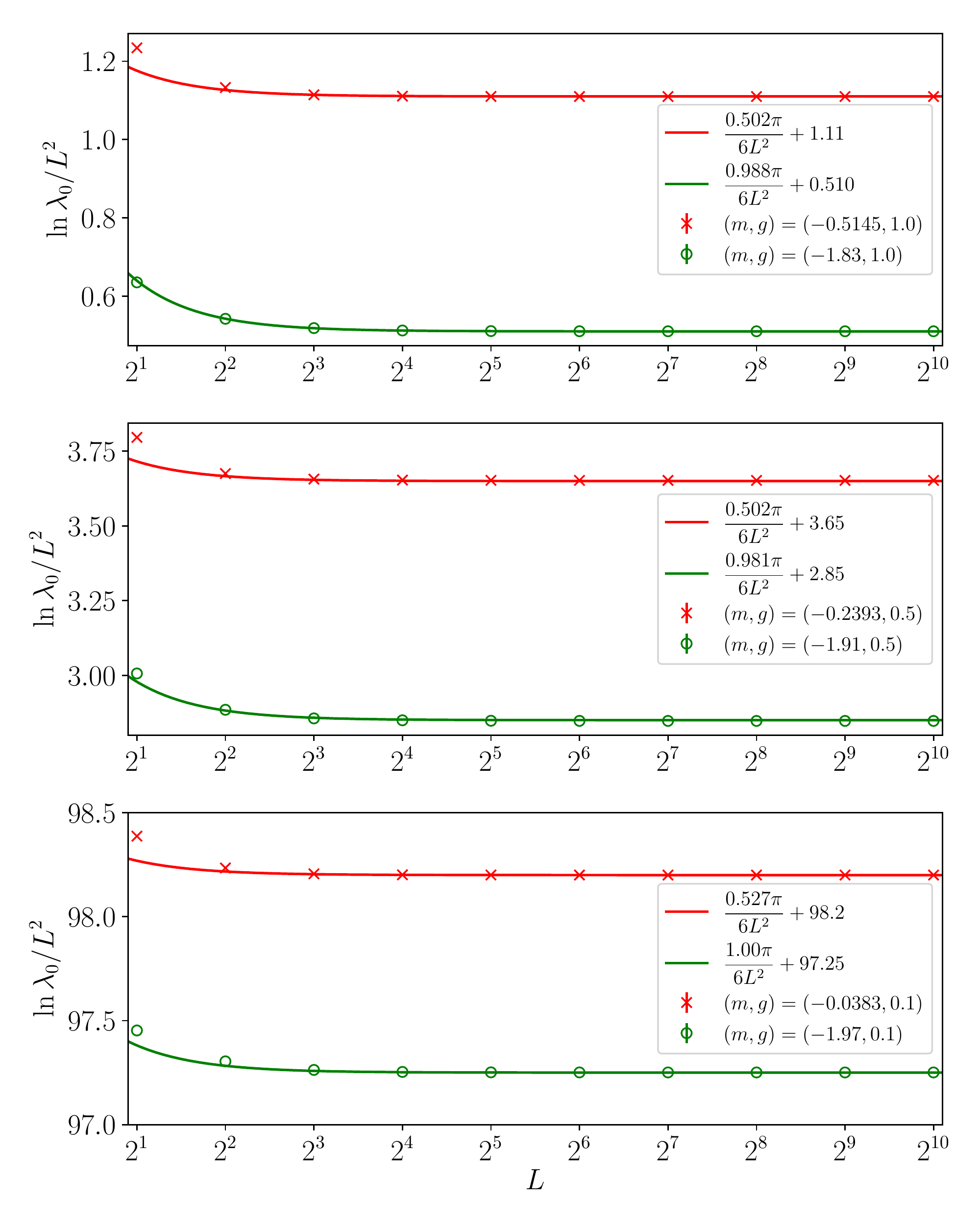}
    \caption{The largest eigenvalue $\lambda_0$ of the density matrix $\rho$ evaluated at
    $g=1.0,\,0.5,\,0.1$ (from top to bottom)  with $D=64$ and $N_\mathrm{ce}=3$ as a function of $L$. Both cases of the
    Ising (red) and BKT (green) transitions are presented. Solid curves represent fit results with
    Eq.~\eqref{eq:l0} for $L\in\left[2^5,2^8\right]$.}
    \label{fig:cc_fc}
  \end{figure}

Carrying out the same analyses as the strong coupling limit, we find two phase transition points at finite
  couplings. Figure~\ref{fig:tp_fc} plots the values of $m_{\rm c}$ at these transition points as a function of $D$, which are determined by the $\langle
  P_\mathrm{odd}\rangle$ analysis. Figure~\ref{fig:cc_fc} shows $L$ dependence of the largest
  eigenvalue of the density matrix $\rho$ at the transition points, whose errors are estimated by the difference between the results with $D=64$ and $32$.
The fit with Eq.~\eqref{eq:l0} gives the central charge listed in Table~\ref{tab:cc_fc}. These results clearly show that two transitions at each coupling
  are the Ising and BKT types and the former moves to $m=0$ and the latter to $m=-2$ toward the weak coupling limit. 

\begin{figure}
    \centering
    \includegraphics[width=80mm]{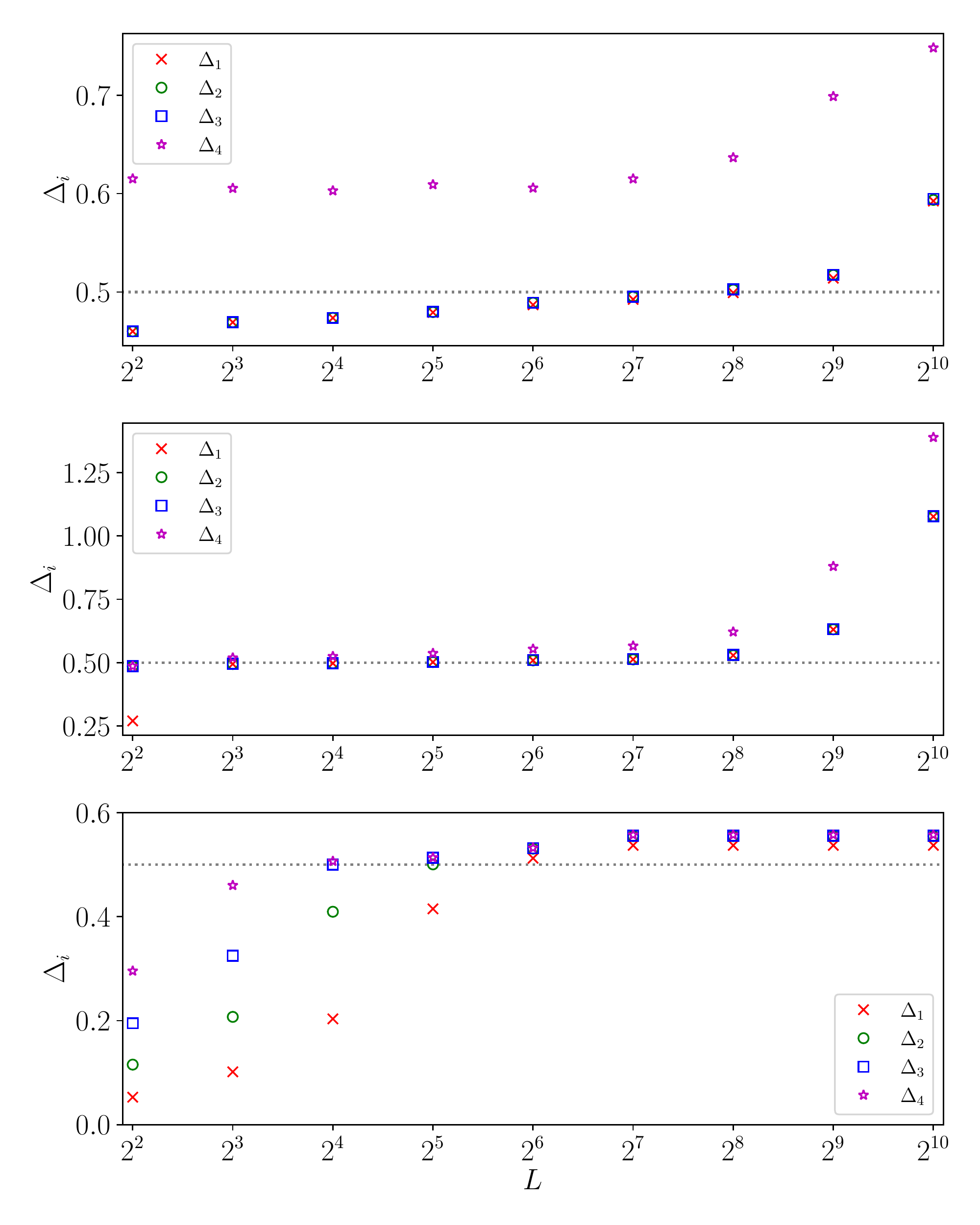}
    \caption{Scaling dimensions $\Delta_1$, $\Delta_2$, $\Delta_3$ and $\Delta_4$ evaluated at
    $(m,g)=(-2,1.0)$ (top), $(-2,0.5)$ (middle) and $(-2,0.1)$ (bottom) with $D=64$ and
    $N_\mathrm{ce}=3$ as a function of $L$. Dotted line denotes $\Delta_i=0.5$ to guide eyes.}
    \label{fig:sd_fc}
  \end{figure}

\begin{figure}
    \centering
    \includegraphics[width=80mm]{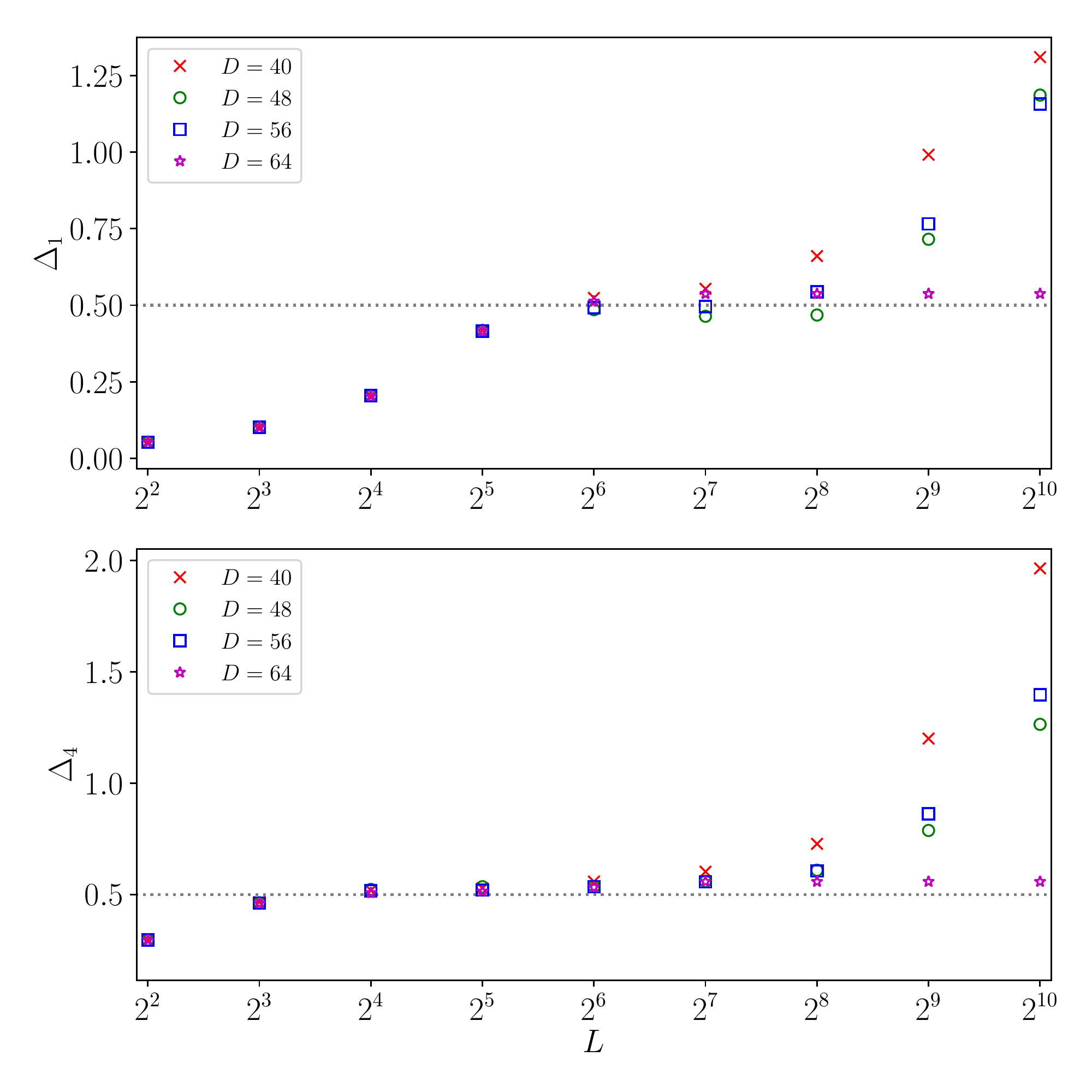}
    \caption{$D$ dependence of scaling dimensions $\Delta_1$ (top) and $\Delta_4$ (bottom) 
evaluated at $(m,g)=(-2,0.1)$ with
    $N_\mathrm{ce}=3$ as a function of $L$. Dotted lines denote $\Delta_1=\Delta_4=0.5$ to guide eyes.}
    \label{fig:dsd_fc}
  \end{figure}

One may expect that the $\mathrm{SU}(2)$ symmetry arises at $m=-2$ even in finite coupling cases. We present our estimate of the scaling dimensions $\Delta_1$, $\Delta_2$, $\Delta_3$ and $\Delta_4$
  at $m=-2$ as a function of $L$ in Fig.~\ref{fig:sd_fc}. The triplet spectrum is observed in all the cases,
  though our results are rather contaminated by the SVD truncation error (see Fig.~\ref{fig:dsd_fc} for $D$
  dependence of the scaling dimensions). 

\begin{figure}[t]
    \centering
    \includegraphics[width=80mm]{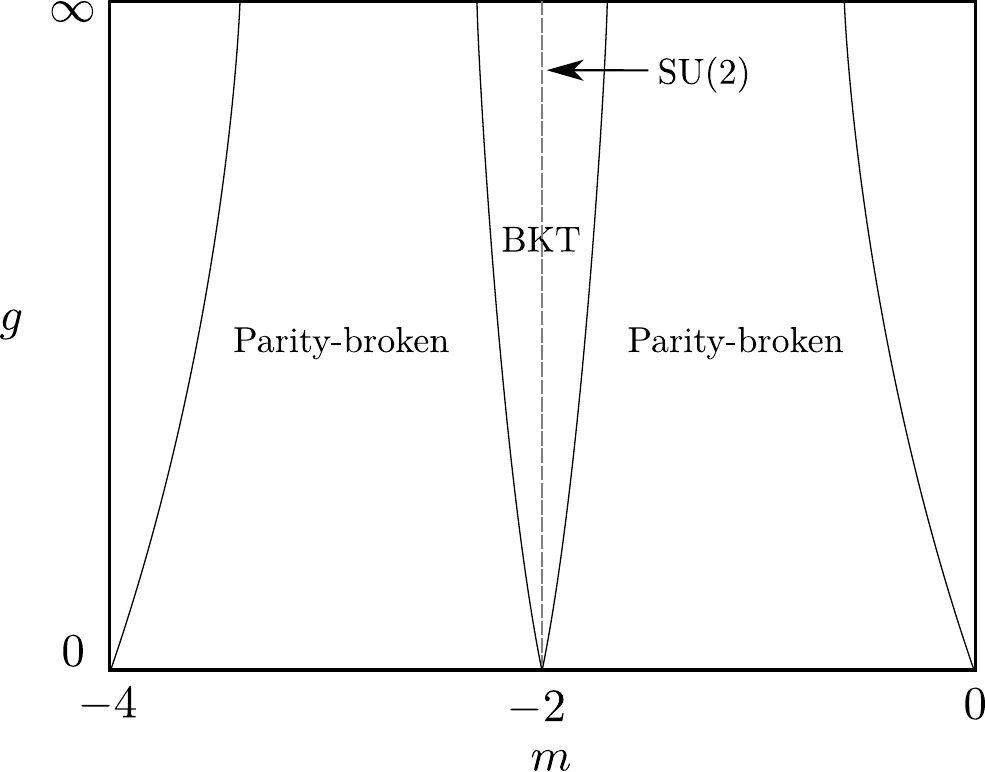}
    \caption{Phase diagram of the lattice Schwinger model with one flavor of the Wilson fermion based on our analyses.}
    \label{fig:pd}
  \end{figure}

From the above investigations, we determine the phase structure of the
  lattice Schwinger model with one flavor of the Wilson fermion as depicted in Fig.~\ref{fig:pd}. The
  reflection symmetry of the Wilson fermion with respect to the $m=-2$ axis is exploited in the figure.
  One of the characteristic features different from the Aoki phase is that only one transition line arises
  from $(m,g)=(0,0)$ (and $(m,g)=(-4,0)$). The same structure is reported in Ref.~\cite{Kenna:2001na} where
  the lattice Gross-Neveu model with one flavor of the Wilson fermion was studied using a method based on the
  weak-coupling expansion.

  \begin{table}[b]
    \centering
    \caption{Fit results of $\lambda_0$ with Eq.~\eqref{eq:l0} at finite couplings for
    $L\in\left[2^5,2^8\right]$. $\lambda_0$ is evaluated with $D=64$ and $N_\mathrm{ce}=3$.}
    \label{tab:cc_fc}
    \begin{ruledtabular}
    \begin{tabular}{ccccc}\toprule
      $g$&$m$&$c$&$f_\infty$&$\sqrt{\chi^2/\mathrm{d.o.f}}$\\ \hline
      $1.0$&$-0.5145$ &$0.50170(22)$&$-1.109792127(56)$&$0.054$ \\
      $1.0$&$-1.83$   &$0.98785(74)$&$-0.51022075(18)$ &$0.0033$ \\
      $0.5$&$-0.2393$ &$0.50169(21)$&$-3.652315026(54)$&$0.037$ \\
      $0.5$&$-1.91$   &$0.9805(25)$ &$-2.84754415(61)$ &$0.0093$ \\
      $0.1$&$-0.0383$ &$0.5270(37)$ &$-98.20031257(84)$&$0.59$ \\
      $0.1$&$-1.97$   &$0.999(24)$  &$-97.2509086(61)$ &$0.052$ \\ \bottomrule
    \end{tabular}
    \end{ruledtabular}
  \end{table}

  \section{Summary and outlook}
  \label{sec:sum}
  We have performed a detailed study of the phase structure of the lattice Schwinger model with one flavor of
  the Wilson fermion by developing a decorated Grassmann TRG method. By implementing an algorithm which
  preserves the reflection symmetry, we
  are allowed to evaluate an expectation value of an projection operator into the parity-odd subspace, which
  reveals a parity-broken phase in a clear manner. We have found not only an Ising phase transition, which has been already known in
  our previous study, but also a BKT phase transition both in the strong coupling limit and the finite coupling region by measuring the
  central charge. Furthermore, our analysis of the scaling dimensions has shown that a CFT with the
  $\mathrm{SU}(2)$ symmetry arises at $(m,g)=(-2,\infty)$, where the lattice
  Schwinger model is known to be equivalent to a critical six-vertex model, and is retained along the line of $m=-2$ in the finite coupling region of the $(m,g)$ plane.

There is a numerical difficulty of accumulation of the SVD truncation error in the study of the BKT phase. The scaling dimensions suffer from a logarithmic finite size correction, which hinders us from quantitative understanding of the properties of the BKT phase at a high level of precision. In recent years, however, the TRG method has been improved rapidly~\cite{Evenbly:2015aaa,Yang:2017aaa,
  Ying:2017aaa,Bal:2017mht,Hauru:2017tne}. These methods allow us to reach larger lattices with higher accuracy
  even at criticality. Further development would open a new path to precise numerical studies 
of the BKT transition.

  It should be noted that the phase structure of the lattice Schwinger model with two or more flavors of the Wilson fermion is also of
  great interest. The Mermin-Wagner theorem prohibits the parity-flavor breaking for finite numbers of
  flavors but massless pseudoscalar mesons appear in the continuum Schwinger model. Extending our
  method to multi-flavor cases is straightforward in principle. However, an efficient way to control the increasing degrees of freedom due to the multi-flavor would be indispensable for practical numerical computation.

  \begin{acknowledgments}
    We thank Shinji Takeda for useful discussions and Hiroshi Ueda for his teaching us the BKT transition.
    Numerical calculations for the present work were mainly carried out using the Cray XC40 system at Yukawa
    Institute for Theoretical Physics, Kyoto University and the HOKUSAI system at Advanced Center for
    Computing and Communication, RIKEN.
    Computation of the preliminary study was performed on the HA-PACS system under the Interdisciplinary Computational Science Program in Center for Computational Sciences, University of Tsukuba. This research is supported by the Ministry of Education, Culture, Sports, Science and Technology (MEXT) as ``Exploratory Challenge on Post-K computer (Frontiers of Basic Science: Challenging the Limits)'' and also in part by Grants-in-Aid for Scientific Research from MEXT (No.~15H03651).
  \end{acknowledgments}

  \appendix
  \section{Detail of the initial tensor}
  \label{app:int}
We list all the nonzero components of the decorated tensor kernel $T^{(b_1,b_2,b_3,b_4)}$ introduced in Eq.~\eqref{eq:tsr}. Here it is useful to classify
$T^{(b_1,b_2,b_3,b_4)}$ into 19 cases according to
  the decoration labels $b_1$, $b_2$, $b_3$ and $b_4$. The following abbreviation is used for simplicity:
  \begin{equation}
    I_{b_1,b_2,b_3,b_4}\equiv\left[I_{b_1}\left(\frac{1}{g}\right)\,
    I_{b_2}\left(\frac{1}{g}\right)\,I_{b_3}\left(\frac{1}{g}\right)\,I_{b_4}\left(\frac{1}{g}\right)
    \right]^{\frac{1}{4}}.
  \end{equation}
  For each tensor index, the value of one (two) corresponds to a parity-even (-odd) state, namely $p(1)=0$
  and $p(2)=1$. Symmetry of the tensor is discussed in Appendix~\ref{app:sym}.

  \subsection{$b_1-b_4=b_3-b_1=b_3-b_2=b_2-b_4=0$}
  The dimension of all tensor indices is two in this case. The nonzero components are as follows:
  \begin{equation}
    \renewcommand{\arraystretch}{1.8}
    \begin{array}{l}
      T^{(b_1,b_2,b_3,b_4)}_{1,1,1,1}=(m+2)^2\,I_{b_1,b_2,b_3,b_4},\\
      T^{(b_1,b_2,b_3,b_4)}_{2,1,2,1}=T^{(b_1,b_2,b_3,b_4)}_{1,2,1,2}=I_{b_1,b_2,b_3,b_4},\\
      \begin{split}
        T^{(b_1,b_2,b_3,b_4)}_{2,2,1,1}&=T^{(b_1,b_2,b_3,b_4)}_{1,2,2,1}=T^{(b_1,b_2,b_3,b_4)}_{1,1,2,2}\\
        &=T^{(b_1,b_2,b_3,b_4)}_{2,1,1,2}=\frac{1}{2}I_{b_1,b_2,b_3,b_4}.
      \end{split}
    \end{array}
    \renewcommand{\arraystretch}{1}
\nonumber
  \end{equation}

  \subsection{$b_1-b_4=b_3-b_2=1$ and $b_3-b_1=b_2-b_4=0$}
  \label{app:int_s2}
  The dimension of the first and third tensor indices is one. That of the second and fourth ones is two.
  The nonzero components are as follows:
  \begin{equation}
    \renewcommand{\arraystretch}{1.8}
    \begin{array}{l}
      T^{(b_1,b_2,b_3,b_4)}_{1,1,1,1}=(m+2)\,I_{b_1,b_2,b_3,b_4},\\
      T^{(b_1,b_2,b_3,b_4)}_{1,2,1,1}=-T^{(b_1,b_2,b_3,b_4)}_{1,1,1,2}=\frac{1}{2}I_{b_1,b_2,b_3,b_4}.
    \end{array}
    \renewcommand{\arraystretch}{1}
\nonumber 
 \end{equation}

  \subsection{$b_1-b_4=b_3-b_2=0$ and $b_3-b_1=b_2-b_4=1$}
  This case is given by a $\frac{\pi}{2}$ rotation of the case~\ref{app:int_s2}.

  \subsection{$b_1-b_4=b_3-b_2=-1$ and $b_3-b_1=b_2-b_4=0$}
  This case is given by a $\pi$ rotation of the case~\ref{app:int_s2}.

  \subsection{$b_1-b_4=b_3-b_2=0$ and $b_3-b_1=b_2-b_4=-1$}
  This case is given by a $-\frac{\pi}{2}$ rotation of the case~\ref{app:int_s2}.

  \subsection{$b_1-b_4=1$, $b_3-b_1=-1$ and $b_3-b_2=b_2-b_4=0$}
  \label{app:int_s3}
  The dimension of the first and second tensor indices is one. That of the third and fourth ones is two.
  The nonzero components are as follows:
  \begin{equation}
    \renewcommand{\arraystretch}{1.8}
    \begin{array}{l}
      T^{(b_1,b_2,b_3,b_4)}_{1,1,1,1}=\frac{m+2}{\sqrt{2}}I_{b_1,b_2,b_3,b_4},\\
      T^{(b_1,b_2,b_3,b_4)}_{1,1,2,1}=T^{(b_1,b_2,b_3,b_4)}_{1,1,1,2}=-\frac{1}{\sqrt{2}}I_{b_1,b_2,b_3,b_4}.
    \end{array}
    \renewcommand{\arraystretch}{1}
\nonumber
  \end{equation}

  \subsection{$b_1-b_4=b_2-b_4=0$ and $b_3-b_1=b_3-b_2=1$}
  This case is given by a $\frac{\pi}{2}$ rotation of the case~\ref{app:int_s3}.

  \subsection{$b_1-b_4=b_3-b_1=0$, $b_3-b_2=-1$ and $b_2-b_4=1$}
  This case is given by a $\pi$ rotation of the case~\ref{app:int_s3}.

  \subsection{$b_1-b_4=b_2-b_4=-1$ and $b_3-b_1=b_3-b_2=0$}
  This case is given by a $-\frac{\pi}{2}$ rotation of the case~\ref{app:int_s3}.

  \subsection{$b_1-b_4=-1$, $b_3-b_1=1$ and $b_3-b_2=b_2-b_4=0$}
  This case is given by a reflection with respect to the $1^\prime$-axis (see Fig.~\ref{fig:ax}) of the
  case~\ref{app:int_s3}.

  \subsection{$b_1-b_4=b_2-b_4=0$ and $b_3-b_1=b_3-b_2=-1$}
  This case is given by a reflection with respect to the $2$-axis (see Fig.~\ref{fig:ax}) of the
  case~\ref{app:int_s3}.

  \subsection{$b_1-b_4=b_3-b_1=0$, $b_3-b_2=1$ and $b_2-b_4=-1$}
  This case is given by a reflection with respect to the $2^\prime$-axis (see Fig.~\ref{fig:ax}) of the
  case~\ref{app:int_s3}.

  \subsection{$b_1-b_4=b_2-b_4=1$ and $b_3-b_1=b_3-b_2=0$}
  This case is given by a reflection with respect to the $1$-axis (see Fig.~\ref{fig:ax}) of the
  case~\ref{app:int_s3}.

  \subsection{$b_1-b_4=b_3-b_1=b_3-b_2=b_2-b_4=1$}
  \label{app:int_s4}
  The dimension of all tensor indices is one in this case. Therefore, the tensor has only one component:
  \begin{equation}
    T^{(b_1,b_2,b_3,b_4)}_{1,1,1,1}=-\frac{1}{2}I_{b_1,b_2,b_3,b_4}.
\nonumber
  \end{equation}

  \subsection{$b_1-b_4=b_3-b_2=-1$ and $b_3-b_1=b_2-b_4=1$}
  This case is given by a $\frac{\pi}{2}$ rotation of the case~\ref{app:int_s4}.

  \subsection{$b_1-b_4=b_3-b_1=b_3-b_2=b_2-b_4=-1$}
  This case is given by a $\pi$ rotation of the case~\ref{app:int_s4}.

  \subsection{$b_1-b_4=b_3-b_2=1$ and $b_3-b_1=b_2-b_4=-1$}
  This case is given by a $-\frac{\pi}{2}$ rotation of the case~\ref{app:int_s4}.

  \subsection{$b_1-b_4=b_2-b_4=1$ and $b_3-b_1=b_3-b_2=-1$}
  \label{app:int_s5}
  The dimension of all tensor indices is one in this case. Therefore, the tensor has only one component:
  \begin{equation}
    T^{(b_1,b_2,b_3,b_4)}_{1,1,1,1}=I_{b_1,b_2,b_3,b_4}.
\nonumber
  \end{equation}

  \subsection{$b_1-b_4=b_2-b_4=-1$ and $b_3-b_1=b_3-b_2=1$}
  This case is given by a $\frac{\pi}{2}$ rotation of the case~\ref{app:int_s5}.

  \section{Rotation and reflection symmetries}
  \label{app:sym}
\begin{figure}
    \centering
    \includegraphics[width=50mm]{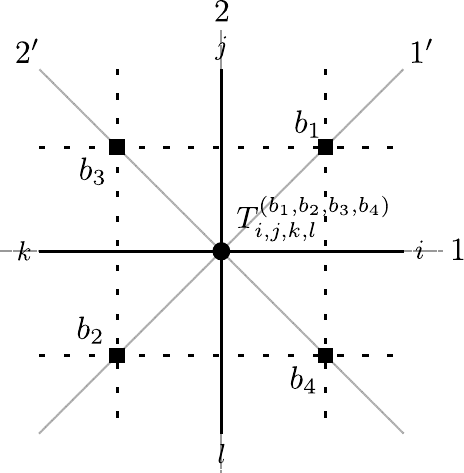}
    \caption{Definition of the $1$-, $2$-, $1^\prime$- and $2^\prime$-axes.}
    \label{fig:ax}
  \end{figure}

  Discrete rotation and reflection symmetries in the lattice action bring some constraints on each
  tensor in the decorated tensor network of Eq.~\eqref{eq:tn}. First, we discuss the discrete rotation symmetry.
  Under a $\frac{\pi}{2}$ rotation, the Grassmann variables $\psi_n$, $\chi_n$ and the decoration labels
  $\left\{b_i\right\}$ transform as
  \begin{gather}
    \begin{split}
      \begin{pmatrix}
        \psi_{n,1}\\\psi_{n,2}
      \end{pmatrix}
      &\to\pm
      \begin{pmatrix}
        \cos\frac{\pi}{4}&\sin\frac{\pi}{4}\\
        -\sin\frac{\pi}{4}&\cos\frac{\pi}{4}
      \end{pmatrix}
      \begin{pmatrix}
        \psi_{n,1}\\\psi_{n,2}
      \end{pmatrix}\\
      &=\pm
      \begin{pmatrix}
        \chi_{n,1}\\-\chi_{n,2}
      \end{pmatrix},
    \end{split}\\
    \begin{pmatrix}
      \chi_{n,1}\\\chi_{n,2}
    \end{pmatrix}
    \to\pm
    \begin{pmatrix}
      \psi_{n,2}\\\psi_{n,1}
    \end{pmatrix},\\
    (b_1,b_2,b_3,b_4)\to(b_4,b_3,b_1,b_2),
  \end{gather}
  where a couple of signs for the Grassmann variables stems from the fact that they transform according to a
  two-valued representation of the rotation group. Then, we get the following constraint by imposing
  the $\frac{\pi}{2}$-rotation symmetry on each tensor:
  \begin{equation}
    \begin{split}
      T^{(b_4,b_3,b_1,b_2)}_{l,i,j,k}=&(-1)^{R(b_4-b_1)+R(b_2-b_3)+b_2+b_4}\\
   &\times T^{(b_1,b_2,b_3,b_4)}_{i,j,k,l}.
    \end{split}
  \end{equation}
  Similarly, other constraints from the $\pi$- and $-\frac{\pi}{2}$-rotation symmetry are obtained:
  \begin{gather}
    \begin{split}
      T^{(b_2,b_1,b_4,b_3)}_{k,l,i,j}=&(-1)^{R(b_4-b_1)+R(b_3-b_1)+R(b_2-b_3)+R(b_2-b_4)}\\
      &\times (-1)^{b_3+b_4}\,T^{(b_1,b_2,b_3,b_4)}_{i,j,k,l},
    \end{split}\\
    \begin{split}
      T^{(b_3,b_4,b_2,b_1)}_{j,k,l,i}=&(-1)^{R(b_1-b_3)+R(b_4-b_2)+b_1+b_4}\\
      &\times T^{(b_1,b_2,b_3,b_4)}_{i,j,k,l}.
    \end{split}
  \end{gather}
  Next, let us turn to the reflection symmetry. Under a reflection with respect to the $1$-axis
  (see Fig.~\ref{fig:ax}), namely space inversion, $\psi_n$, $\chi_n$ and $\left\{ b_i \right\}$ transform as
  \begin{gather}
      \begin{pmatrix}
        \psi_{n,1}\\\psi_{n,2}
      \end{pmatrix}
      \to\gamma_1
      \begin{pmatrix}
        \psi_{n,1}\\\psi_{n,2}
      \end{pmatrix}
      =
      \begin{pmatrix}
        \psi_{n,1}\\-\psi_{n,2}
      \end{pmatrix},\\
    \begin{pmatrix}
      \chi_{n,1}\\\chi_{n,2}
    \end{pmatrix}
    \to
    \begin{pmatrix}
      \chi_{n,2}\\\chi_{n,1}
    \end{pmatrix},\\
    (b_1,b_2,b_3,b_4)\to(-b_4,-b_3,-b_2,-b_1).
  \end{gather}
  Symmetry under these transformations brings
  \begin{equation}
    \begin{split}
      T^{(-b_4,-b_3,-b_2,-b_1)}_{i,l,k,j}=&(-1)^{R(b_4-b_1)+R(b_2-b_3)}\\
      &\times (-1)^{(b_1+b_3)(b_2+b_3)+(b_1+b_2)(b_2+b_4)}\\
      &\times (-1)^{p(i)+p(j)+p(k)+p(l)}\,T^{(b_1,b_2,b_3,b_4)}_{i,j,k,l}.
    \end{split}
  \end{equation}
  Transformation under a reflection with respect to the $2$-axis (see Fig.~\ref{fig:ax}), namely time
  inversion is analogous:
  \begin{gather}
      \begin{pmatrix}
        \psi_{n,1}\\\psi_{n,2}
      \end{pmatrix}
      \to\gamma_2
      \begin{pmatrix}
        \psi_{n,1}\\\psi_{n,2}
      \end{pmatrix}
      =
      \begin{pmatrix}
        \psi_{n,2}\\\psi_{n,1}
      \end{pmatrix},\\
    \begin{pmatrix}
      \chi_{n,1}\\\chi_{n,2}
    \end{pmatrix}
    \to
    \begin{pmatrix}
      \chi_{n,1}\\-\chi_{n,2}
    \end{pmatrix},\\
    (b_1,b_2,b_3,b_4)\to(-b_3,-b_4,-b_1,-b_2),
  \end{gather}
  and a derived constraint is
  \begin{equation}
    \begin{split}
      T^{(-b_3,-b_4,-b_1,-b_2)}_{k,j,i,l}=&(-1)^{R(b_1-b_3)+R(b_4-b_2)}\\
      &\times (-1)^{(b_1+b_3)(b_1+b_4)+(b_2+b_3)(b_3+b_4)}\\
      &\times (-1)^{p(i)+p(j)+p(k)+p(l)}\,T^{(b_1,b_2,b_3,b_4)}_{i,j,k,l}.
    \end{split}
  \end{equation}
  We consider reflections with respect to the $1^\prime$- and $2^\prime$-axes also which are defined in
  Fig.~\ref{fig:ax}. That with respect to the $1^\prime$-axis is interpreted as a $\frac{\pi}{2}$ rotation
  followed by a reflection with respect to the $1$-axis. The other is a -$\frac{\pi}{2}$ rotation followed by
  a reflection with respect to the $1$-axis. Thus, the following constraints are obtained:
  \begin{gather}
    \begin{split}
      &T^{(-b_1,-b_2,-b_4,-b_3)}_{j,i,l,k}\\
=&(-1)^{(b_1+b_3)(b_1+b_4)+(b_2+b_3)(b_2+b_4)}\\
      &\times (-1)^{p(i)+p(j)+p(k)+p(l)}\,T^{(b_1,b_2,b_3,b_4)}_{i,j,k,l},
    \end{split}\\
    \begin{split}
      &T^{(-b_2,-b_1,-b_3,-b_4)}_{l,k,j,i}\\
=&(-1)^{R(b_1-b_4)+R(b_1-b_3)+R(b_3-b_2)+R(b_4-b_2)}\\
      &\times (-1)^{(b_1+b_3)(b_1+b_4)+(b_2+b_3)(b_2+b_4)+b_3+b_4}\\
      &\times (-1)^{p(i)+p(j)+p(k)+p(l)}\,T^{(b_1,b_2,b_3,b_4)}_{i,j,k,l}.
    \end{split}
  \end{gather}

%

\end{document}